\documentclass[%
 reprint,
superscriptaddress,
 amsmath,amssymb,
 aps, pre,
floatfix
]{revtex4-1}

\usepackage{bm}
\usepackage[version=4]{mhchem}
\usepackage{siunitx}
\usepackage[format=plain,indention=0cm,justification=raggedright,font=footnotesize]{caption}
\usepackage{graphicx}
\usepackage{bm}
\usepackage{dcolumn}

\DeclareMathOperator{\erf}{erf}

\newcommand{\beginsupplement}{%
        \setcounter{table}{0}
        \renewcommand{\thetable}{S\arabic{table}}%
        \setcounter{figure}{0}
        \renewcommand{\thefigure}{S\arabic{figure}}%
        \setcounter{equation}{0}
        \renewcommand{\theequation}{S\arabic{equation}}%
     }

\newcommand{\figref}[1]{Fig.~\ref{#1}}
\newcommand{\secref}[1]{Sec.~\ref{#1}}
\newcommand{\eqnref}[1]{Eq.~\eqref{#1}}
\renewcommand\vec{\mathbf}

\begin{document}

\title{A fine balance of chemotactic and  hydrodynamic torques: when microswimmers orbit a pillar just once}

\author{Chenyu Jin}
\affiliation{Max Planck Institute for Dynamics and Self-Organization, Am Fa\ss{}berg 17, 37077 G\"ottingen, Germany.}

\author{J\'er\'emy Vachier}
\affiliation{Max Planck Institute for Dynamics and Self-Organization, Am Fa\ss{}berg 17, 37077 G\"ottingen, Germany.} 

\author{Soumya Bandyopadhyay}
\affiliation{School of Mechanical Engineering, Purdue University, West Lafayette, Indiana 47907, USA.}

\author{Tamara Macharashvili}
\affiliation{Princeton University, New Jersey, NJ 08544, USA. }

\author{Corinna C. Maass}
\email{corinna.maass@ds.mpg.de}
\affiliation{Max Planck Institute for Dynamics and Self-Organization, Am Fa\ss{}berg 17, 37077 G\"ottingen, Germany.}

\date{\today}
\begin{abstract}
We study the detention statistics of self-propelling droplet microswimmers attaching to microfluidic pillars. These droplets show negative autochemotaxis: they shed a persistent repulsive trail of spent fuel that biases them to detach from pillars in a specific size range after orbiting them just once. 
We have designed a microfluidic assay recording swimmers in pillar arrays of varying diameter, derived detention statistics via digital image analysis and interpreted these statistics via the Langevin dynamics of an active Brownian particle model.
By comparing data from orbits with and without residual chemical field, we can independently estimate quantities like hydrodynamic and chemorepulsive torques, chemical coupling constants and diffusion coefficients, as well as their dependence on environmental factors like wall curvature.
\end{abstract}
\maketitle
Biological microswimmers operate in complex geometries and react to chemical and physical gradients, both external, e. g.\ in nutrients (chemotaxis~\cite{madigan2005_bacterial, adler1966_chemotaxis,hazelbauer2012_bacterial}), and self generated ones (autochemotaxis~\cite{bonner1947_evidence,zhao2013_psl,reid2012_slime}), the latter enabling them to communicate and cooperate.
The boundary conditions imposed e.g.\ by soil packings and interfaces affect both the active dynamics of swimmers, via hydrodynamic wall interactions~\cite{spagnolie2012_hydrodynamics, berke2008_hydrodynamic} or active accumulation~\cite{li2009_accumulation, vanteeffelen2008_dynamics, elgeti2013_wall}, as well as the diffusive spread of chemotactic fields. 
The coupling of these dynamics leads to complex feedback phenomena that can significantly influence behaviour like arrest, aggregation and biofilm formation~\cite{tuval2005_bacterial}.

In order to derive and test tractable models for systems featuring such complex feedback it is important to develop experimental assays to decouple and control individual contributions.

We have previously demonstrated one quite unexpected phenomenon in a highly tunable artificial model system, while studying the interaction of self-propelling droplet swimmers with pillars in quasi-2D microfluidic cells. These droplets shed a chemorepellent that biases them towards detachment after circling the pillar once~\cite{jin2018_chemotactic}. Our system has the advantage of directly comparing wall attachment with and without chemotactic repulsion, as well as controllable curvature set by the pillar size.

In this paper we present a quantitative study of the pillar interaction, analyzing the data from multiple experimental data sets in the context of an analytical Langevin model incorporating hydrodynamic, chemotactic and stochastic torques.
From our quantitative analysis we gain insight into the respective strengths of wall attraction, chemorepulsion, as well as rotational diffusion and their effects on interfacial capture.

Our swimmers consist of oil droplets of radius $a\approx \SI{50}{\um}$,  dissolving gradually in a micellar aqueous surfactant solution, on a time scale of up to several hours (See SI \secref{SIsec:methods} for experimental protocols). 
The oil solubilises by diffusing into surfactant micelles in a boundary layer around the droplet, a process that also depletes the surfactant coverage of the droplet's oil-water interface~\cite{herminghaus2014_interfacial,maass2016_swimming}. 
If the droplet position fluctuates, the symmetric cloud of filled micelles is deformed towards trailing behind the droplet, such that there are more empty micelles at the anterior. Empty micelles can disintegrate to replenish the surfactant coverage at the interface: thus, a moving droplet has more surfactant at its anterior, leading to a self-sustaining gradient in interfacial tension towards the posterior and a force propelling the droplet ($v\approx$\,30--60\SI{}{\um\per\second}). 
\begin{figure}
    \centering
    \includegraphics[width=\columnwidth]{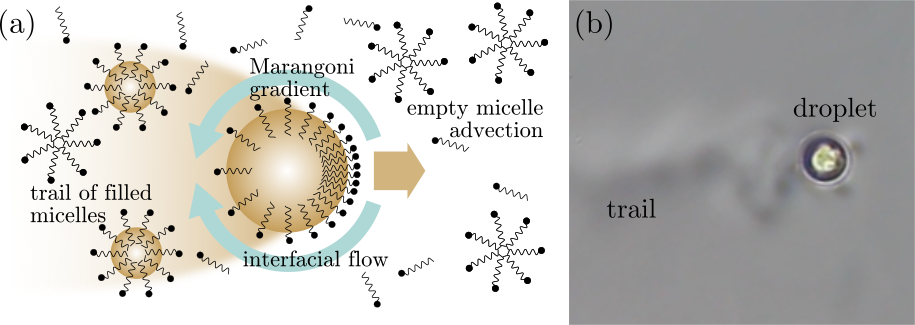}
    \caption{(a) Schematic of the the droplet propulsion via interfacial Marangoni gradients. (b) Phase contrast image of a swimming droplet, dragging a trail of filled micelles visible due to refractive index variation. Figure adapted from~\cite{jin2017_chemotaxis}. }
    \label{fig:propulsion}
\end{figure}

Chemotaxis and autochemotaxis follow naturally from the droplets' mechanism of propulsion~\cite{jin2017_chemotaxis}.
Droplets are attracted by empty micelles, such that they follow surfactant gradients (chemotaxis) and avoid areas of filled micelles, which act as a chemorepellent. 
The micelles diffuse slowly in comparison to the droplet motion, with a diffusion constant $D_\text{m}\approx\,\SI{100}{\micro\metre\squared\per \second}$~\cite{candau1984_new}, such that the repulsive trail of filled micelles persists over cruising ranges of tens to hundreds of droplet diameters.
We have experimentally confirmed~\cite{jin2017_chemotaxis} that this negative autochemotaxis is indeed mediated by the diffusion of filled micelles. 

\begin{figure}
\centering
  \includegraphics[width=.8\columnwidth]{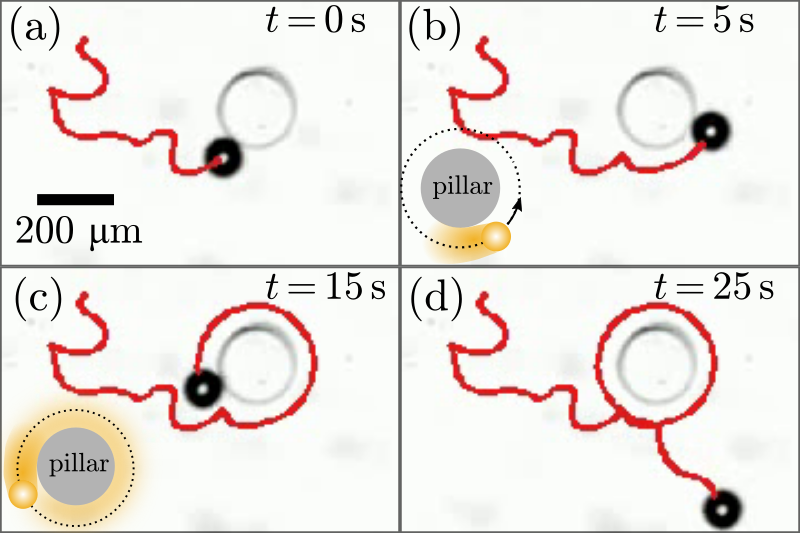}
  \caption{A negatively autochemotactic swimmer ($a=$\SI{50}{\um}) interacting with a pillar ($R=$\SI{100}{\um}). (a) The swimmer attaches to the pillar, (b, c) travels around the pillar and (d) detaches soon after it re-enters its own trail. The trajectory of the swimmer is plotted onto the optic microscopic images. The inserts illustrate the trail behind the swimmer.}
  \label{fig:trail}
\end{figure}

\begin{figure}
\centering
  \includegraphics[width=.8\columnwidth]{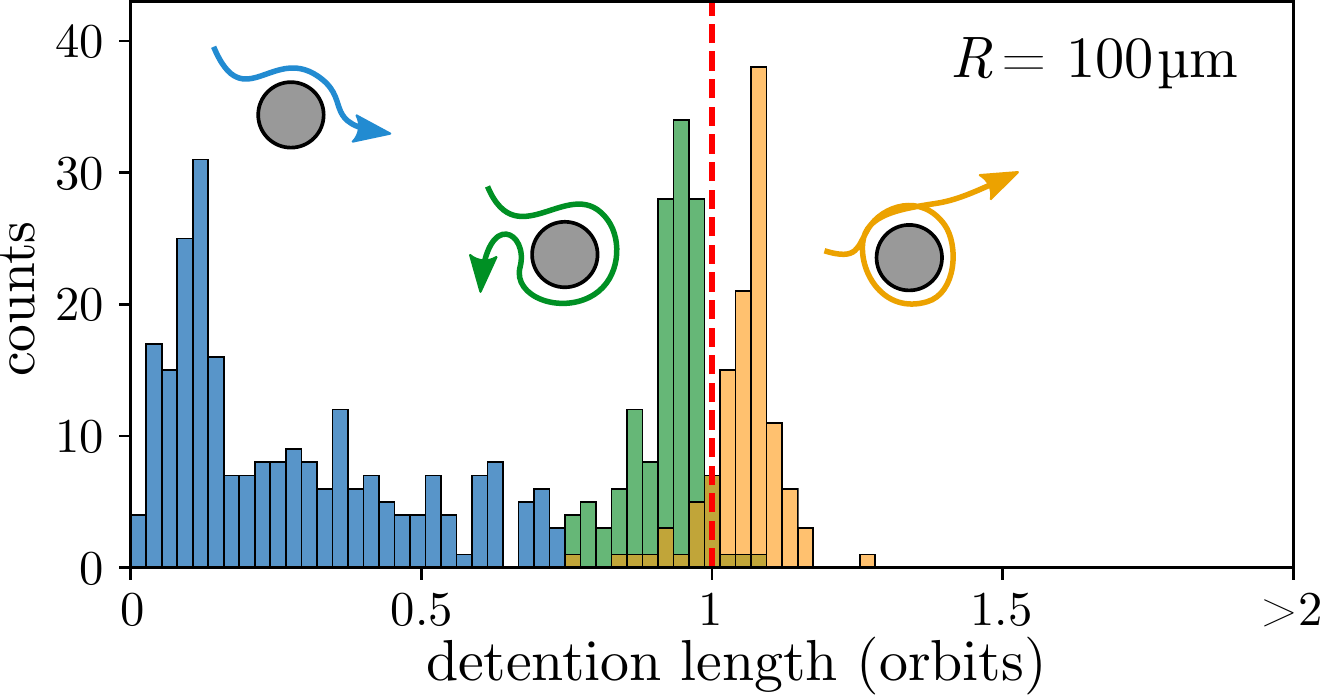}
  \caption{Example histogram of the detention of swimmers ($a=$\SI{50}{\um}) at a pillar ($R=$\SI{100}{\um}). Colour codes and sketches illustrate the cases of scattering (blue), non-crossing (green) and self-crossing (orange) trajectories. Note that we bin by length here for illustration, but will use detention times in the remainder to be consistent with theory.}
  \label{fig:sample-hist}
\end{figure}

\begin{figure}
  \centering
  \includegraphics[width=.8\columnwidth]{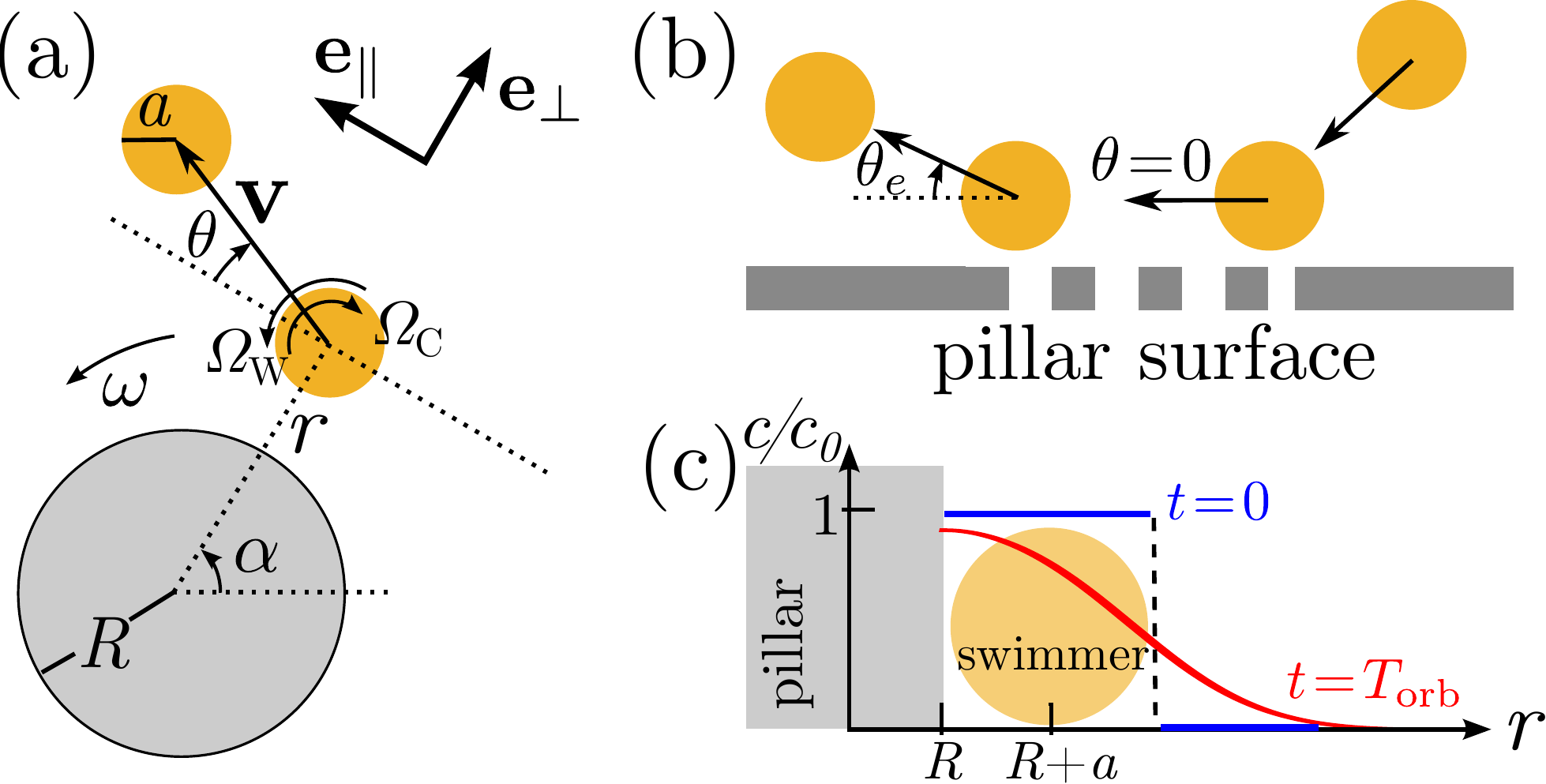}
  \caption{Model schematics and coordinate definitions. (a) a swimmer of radius $a$ moving around a pillar of radius $R$ with velocity $\mathbf{v}$ and angular velocity $\omega$, torques from chemotactic, $\Omega_\text{C}$, and wall interaction $\Omega_\text{W}$ and polar coordinates ($r$, $\alpha$). (b) Schematic drawing of a swimmer approaching and leaving a pillar, with escape angle $\theta_\text{e}$. (c) Chemorepellent diffusion away from the pillar (numerical solution of \eqnref{eqn:main} for $R=a$). Radial concentration $c/c_0$ vs. $r$, with the initial step function (dotted, blue) for $t=0$ and the diffused profile (solid, red) for $t=T_\text{orb}$.}
  \label{fig:scheme}
\end{figure}

A persistent swimmer will not successfully attach to a convex wall without an attractive torque, simply for reasons of geometry. Since the required torque increases with curvature, swimmers experiencing a finite wall attraction are expected to scatter off small pillars and get trapped at large ones~\cite{sipos2015_hydrodynamic, takagi2014_hydrodynamic,simmchen2016_topographical}. Our specific swimmers are attracted to walls and indeed exhibit scattering from small and trapping at large pillars. 
However, at pillars of intermediate radii, we observe a significant increase of detachment after circling the pillar once, i.e.\ when swimmers approach and re-enter their own trajectory \cite{jin2018_chemotactic}: the filled micelles diffuse away from the pillar (\figref{fig:scheme}\,(c)), providing an additional chemorepulsive gradient and favouring detachment. 
We illustrate this by consecutive video stills of a typical interaction with a \SI{100}{\micro \metre} pillar in \figref{fig:trail} and a histogram of detention lengths for many similar interactions in \figref{fig:sample-hist}, which clearly shows a peak around lengths corresponding to one orbit (see also supporting movie, \secref{SIsec:movie}).

The long-tailed peaks in the statistics of detention times suggest stochastic influences on the pillar interaction \cite{schaar2015_detention, spagnolie2015_geometric}.
Rotational noise and convex curvature cause a fraction of swimmers to detach almost immediately (\figref{fig:sample-hist}, blue peak). 
For droplets attached long enough to approach their own residual chemical field, we can distinguish two cases: 
If the chemical repulsion is too strong to allow the droplets to re-enter their trajectory, this creates a \textit{non-crossing} peak in the histogram of detention lengths slightly before one full orbit (green).
Otherwise, the droplet is forced to swim along its own trail, as traced in \figref{fig:trail}. However, the added chemotactic repulsion adds a bias towards detachment, resulting in a \textit{self-crossing} peak in the histogram of detention lengths slightly after one orbit (yellow).
We note that we exclude data from non-crossing interactions from our analysis, since their chemical interactions break radial symmetry and would require rather sophisticated modelling. We can already estimate the forces and torques of interest from the following simple analytical model applied to the cases of short-time scattering and self-crossing.

For a quantitative estimate of the effects of curvature, wall interaction and autochemotaxis we model our swimmers as active Brownian particles (ABP)~\cite{romanczuk2012_active,schmitt2016_active}.
The swimmer is represented by a point particle moving with speed $v$ in the direction $\mathbf{e}$. 
The dynamics of position $\mathbf{r}$ and direction $\mathbf{e}$ are described by the overdamped Langevin equations
\begin{equation}
\begin{aligned}
  \dot{\vec{r}} &= v \vec{e}\\
  \dot{\vec{e}} &= (\boldsymbol{\Omega} + \sqrt{2D_\text{R}} \boldsymbol{\xi}) \times \vec{e}.
\end{aligned}
\label{eqn:langevin} 
\end{equation}
$\boldsymbol{\Omega}$ is the sum of deterministic drifts of $\mathbf{e}$, and the rotational fluctuations are modeled by the rotational diffusivity $D_\text{R}$ and a normalized zero-mean Gaussian white noise $\boldsymbol{\xi}$ with variance $\langle\boldsymbol{\xi}_i(t)\boldsymbol{\xi}_j(t') \rangle = \delta_{ij}\delta(t-t')$.

We note that in active emulsions, the orientation can be affected by both whole body rotations and shifts in interfacial tension. However, in our reduction to ABP point particles, these revert to effective forces and torques.

The droplet motion is restricted to the $xy$ plane, with a perpendicular $\boldsymbol{\Omega} = (0, 0, \Omega)^T$. 
In our system $\Omega = \Omega_\text{W}+\Omega_\text{C}$ comprises the sum of deterministic drifts of $\mathbf{e}$: the external (mostly hydrodynamic) torque due to wall attraction, $\Omega_\text{W}$, and autochemotaxis, $\Omega_\text{C}$, from the second orbit.
A schematic representation of the system is shown in \figref{fig:scheme}(a).
We use polar coordinates $(r,\alpha)$ with the origin $r=0$ at the center of a pillar of radius $R$, with a swimmer of radius $a$ at the position $\vec{r}$ from the origin. We project the orientational unit vector of the swimmer $\mathbf{e}$ on two orthogonal unit vectors, a radial one pointing towards the pillar center, $\mathbf{e}_\perp$,
and one tangential to the pillar interface, $\mathbf{e}_\parallel$.
We define $\theta \in [-\pi/2,\pi/2]$ as the angle between $\mathbf{e}$ and $\mathbf{e}_\parallel$, with $\theta \approx 0$ when the droplet is moving along the pillar tangent.
Without loss of generality, we assume the swimmer to move counterclockwise around the pillar, such that it approaches with $\theta >0$ and detaches with $\theta <0$.
From \eqnref{eqn:langevin}, we derive \cite{spagnolie2015_geometric}
\begin{equation}
\begin{aligned}
  \dot{r} &= \dot{\vec{r}} \cdot \vec{e}_\perp  = -v \sin\theta  \\
  \dot{\theta} &= -\omega + \Omega + \sqrt{2D_\text{R}} \xi,\text{ with } \omega=\frac{1}{r} (\dot{\mathbf{r}} \cdot \vec{e}_\parallel),
\end{aligned}
\label{eqn:langevin-theta}
\end{equation}
where $\omega>0$ is the angular velocity of the swimmer around the pillar.
The swimmer detaches when $\theta$, starting from $\theta_0$, exceeds a fixed escape angle $\theta_\text{e}$~\cite{zottl2016_emergent,spagnolie2015_geometric} (\figref{fig:scheme}(b)).
The detention time $t_\text{d}$ is therefore the first-passage time of $\theta_\text{e}$ of a stochastic process described by \eqnref{eqn:langevin-theta}.
Here we take $\theta_0 = 0$, and $\theta_\text{e} = -0.962$ as the escape angle measured at the moment of detachment (see SI \secref{SIsec:criteria}).

Our restriction to counterclockwise motion would in principle necessitate a reflective boundary condition for $\theta_r=\pi/2$, however, for moderate rotational diffusion $D_r$ (see also SI~\secref{SIsec:diffusion} and \ref{SIsec:FP}), this head-on orientation is hydrodynamically unlikely in the case of our highly persistent pusher-type swimmers~\cite{herminghaus2014_interfacial, kruger2016_curling} even in the limiting case of zero wall curvature and no chemotactic wall repulsion, and the boundary condition can be safely neglected.

Generally, $\Omega_\text{W}$ and $\Omega_\text{C}$ are functions of $r(t),\,\theta(t)$ and $\alpha(t)$, depending on the exact nature of the hydrodynamic and chemical fields.
However, we can treat them as constants using the following approximations. We assume the transition between attached and detached state to be instantaneous compared to the detention time, such that $r \approx R+a$, $\theta\approx0$ and $\Omega_\text{W}$ is constant for the swimmer while it is attached. We note that we cannot expect $\Omega_\text{W}$ to be the same for all pillar sizes, if the length scales of pillar, droplet and flow field,  e.g.\ puller dipole size, are comparable (cf.~\cite{spagnolie2012_hydrodynamics}).

The recorded angular speeds $\omega$ are quite uniform in each orbit (see SI \figref{SIfig:omega}), with a slight slowdown in the second orbit, when the droplet moves on its own trail. 
We therefore treat $\omega$ as constant within one orbit, with $\omega\in {\omega_1,\omega_2}$. 

For $(\omega_1-\omega_2)/\omega_1\ll1$ (cf.~\figref{fig:analysis} (a)), we can further assume that the orbiting period $T_\text{orb}$ of the droplet does not significantly increase with time, such that the droplet always experiences the same chemical gradient $\partial_r c(t)|_{r=R+a}$ initiated during its previous passage. 
We therefore approximate $\Omega_\text{C}$ as constant between the points where the droplet has fully crossed over onto its trail and where it leaves the pillar, and we define the detention time $t_\text{d}-T_\text{orb}$ in the second orbit as the time elapsed between these two events.

Under these approximations of constant $\Omega_\text{W}$, $\Omega_\text{C}$ and $\omega$, the detention time $t_\text{d}$ follows an inverse Gaussian distribution
\begin{equation}
  f(t_\text{d}; \mu, \lambda) = \sqrt{ \frac{\lambda}{2 \pi t_\text{d}^3} } \exp \left[ \frac{-\lambda (t_\text{d}-\mu)^2}{2\mu^2t_\text{d}} \right],
  \label{eqn:IG}
\end{equation}
with $\lambda = \theta_\text{e}^2/2D_\text{R}$ and the distribution mean $\mu = \langle t_\text{d} \rangle = \theta_\text{e} / (\Omega-\omega)$.

If the deterministic torque compensates the geometry effect, the drift $\omega + \Omega$ is zero, and $\mu$ diverges. In this case, $f(t_\text{d})$  reverts to a L\'evy distribution
\begin{equation}
    f(t_d)=\sqrt{ \frac{\lambda}{2 \pi t_\text{d}^3} } \exp \left[ \frac{-\lambda}{2t_\text{d}} \right]\,.
    \label{eqn:Levy}
\end{equation}

We have tested our approximations for the case of a straight wall without drift by calculating  $f(t_d)$ using a less approximative Fokker-Planck model~\cite{schaar2015_detention} that includes the reflective boundary condition. The difference to the corresponding  L\'evy distribution is indeed too small to be resolved in experimental statistics (see SI,~\secref{SIsec:FP}).

We numerically approximate the concentration field $c(r, \alpha)$ of filled micelles in the polar coordinate system $(r,\alpha$) (\figref{fig:scheme}). 
We assume that a droplet of radius $a$ dissolves at a constant rate and approximate its initial trail of filled micelles by a step function, $c(R<r<R+2a) = c_0$,  which we let diffuse over time, using the micellar diffusion coefficient $D_\mathrm{m} \approx \SI{100}{\um\squared\per\second}$. We further assume the diffusion coefficients of filled and empty micelles to be similar. 

Under the boundary conditions $\partial_r c|_{r=R} = 0$ and $c(r\rightarrow +\infty) =0$
we solve the diffusion equation
\begin{equation}
\partial_t c = D_\text{m} \partial_r\left( r \partial_r c\right) /r,
  \label{eqn:main}
\end{equation}
 using a forward-marched explicit finite difference scheme.
The concentration profile is computed numerically with Matlab, selecting an appropriate grid size based on the stability criteria for one-dimensional parabolic partial differential equations \cite{grundmann2009_boundary}.

For illustration, we plot in \figref{fig:scheme}(c) the rescaled filled micelle concentrations $c/c_0$ behind a swimmer attached to a pillar with $R=a$, diffusing from a step function at $t=0$ (dashed, blue) to a smoothed profile (solid, red) at $t=T_\text{orb}$. Calculated profiles and derivatives for our experimental settings are listed in the SI,  \secref{SIsec:gradient}.

For a statistical analysis of detention times, we recorded multiple sets of small numbers
of monodisperse droplets of radius  $a=\SI{50}{\micro\metre}$ interacting with microfluidic pillar arrays of radius $R\in\{50,75,100,250\}\SI{}{\micro\metre}$. Data sets for even larger pillar sizes are included in SI~\secref{SIsec:histograms}, however, since droplets are effectively trapped at these pillars, we are not using them for detention statistics.

By restricting the analysis to ``clean'' pillars without residual chemorepellent from previous interactions, we retain several hundred interactions per pillar size.

\begin{figure}
 \centering
 \includegraphics[width=\columnwidth]{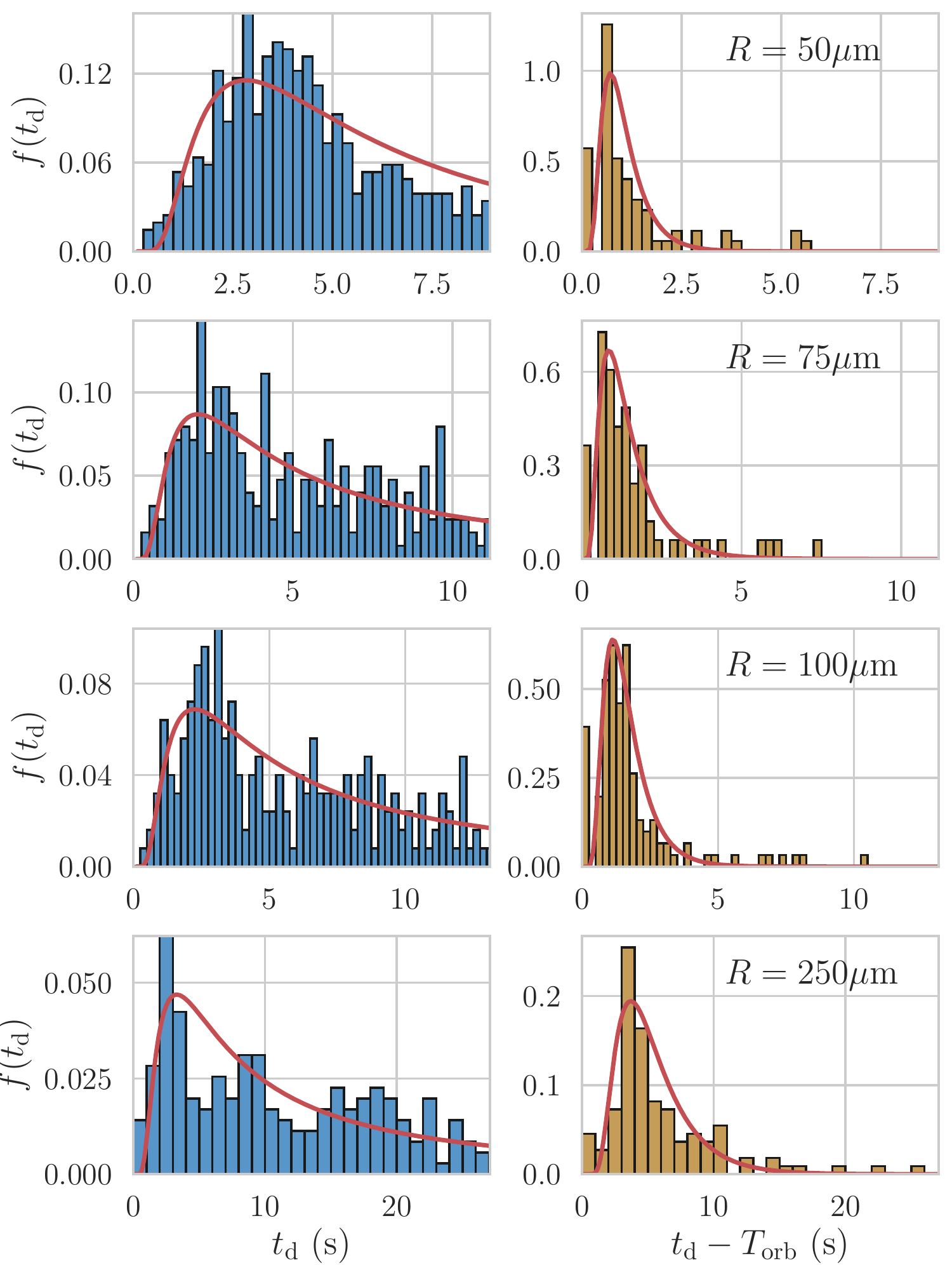}
 \caption{Normalized histograms $f(t_d)$ of the detention times of $a = $\SI{50}{\um} swimmers at pillars of varying radii $R$, first (blue, left) and second, self-crossing, (yellow, left) orbits. The $t$ origin of the latter is reset for each trajectory to the point $T_\text{orb}$ where the droplet crosses its own trajectory. Red lines: inverse Gaussian fits, first orbit fitted only for $t<0.7 T_\text{orb}$ to rule out non-crossing chemotactic detachments. 
 }
 \label{fig:hist-fit}
\end{figure}

\begin{figure*}
    \centering
    \includegraphics[width=\textwidth]{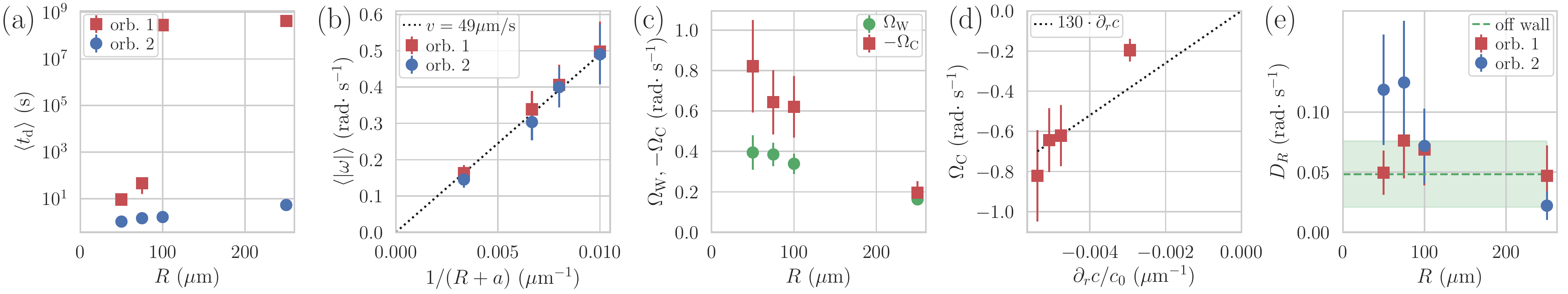}
    \caption{Extracted parameters for detention statistics of $a=\SI{50}{\micro\metre}$ at pillars with $R\in\{50,75,100,250\}\SI{}{\micro\metre}$, values for first and second orbit. (a) measured detention times. (b) measured orbital speed vs. curvature. (c) wall attraction $\Omega_\text{W}$ and negative chemorepulsion  $-\Omega_\text{C}$ torques, from fits. (d) $\Omega_\text{C}$ vs. calculated chemical gradient, (e) rotational diffusion coefficient, from fits, green dashed line and shaded confidence interval: free motion between pillars. For numerical values and error estimates see tables in SI, section~\ref{SIsec:analysis}.}
    \label{fig:analysis}
\end{figure*}

We have sorted all attachment events into scattering, self-crossing and non-crossing events (\figref{fig:sample-hist}, see SI for image analysis criteria and statistical quantities) and extracted the quantities $t_\text{d}, v(t)$, $\omega(t)$ and $\theta(t)$.  
\figref{fig:hist-fit} contains two subsets of detention time distributions for increasing pillar sizes and the respective inverse Gaussian fits of \eqnref{eqn:IG}. The first set (blue) contains the data range for $0<t_\text{d}<0.7\, T_\text{orb}$, during which time we expect no chemical field, $\Omega_\text{C}=0$.
The second set (yellow, only self-crossing) contains detention times taken from the moment $T_\text{orb}$ where the droplet crosses its own trajectory and $\Omega_\text{C}=\text{const}$.

Using the extracted fitting parameters $\mu, \lambda$ from \eqnref{eqn:IG} and the measured $\omega$, we calculate $D_\text{R}, \Omega_\text{W}$ and $\Omega_\text{C}$ for each pillar size, as well as the expectation value of the detention times $\langle t_\text{d} \rangle =\mu$, as plotted in \figref{fig:analysis}.

As expected,  $\langle t_\text{d} \rangle$ is larger in the first orbit (\figref{fig:analysis} (a)), where $\Omega_\text{C}=0$, and increases with increasing $R$, in fact diverging for $\Omega_\text{C}=0$ and $R\in\{100,250\}\SI{}{\micro\metre}$, indicating a trapped state.
The recorded mean orbital speeds $\langle|\omega|\rangle$ scale linearly with $(R+a)^{-1}$ and do not change significantly between orbits 1 and 2, justifying our approximation of an overall constant speed $v\approx\SI{47}{\micro\metre\per\second}$ (\figref{fig:analysis} (b)).

The extracted torques (\figref{fig:analysis} (c)) allow for a number of observations: both $\Omega_\text{W}$ and $-\Omega_\text{C}$ decrease with pillar size, the latter simply because orbiting takes longer and the chemorepulsive gradient $\partial_r c$ decreases by diffusion, the former possibly because the pusher type hydrodynamic interactions depend on wall curvature~\cite{spagnolie2012_hydrodynamics}.
$\Omega_\text{C}$ overcompensates $\Omega_\text{W}$ in all cases up to $R=\SI{250}{\micro\metre}$, explaining the enhanced detachment after one orbit and the transition to fully trapped states for $R>\SI{250}{\micro\metre}$. Similarly, the diverging detention times in the first orbit for $R\in\{100,250\}\SI{}{\micro\metre}$ are due to the fact that $\Omega_\text{W}$ compensates $\omega$ while $\Omega_\text{C}=0$.

For an estimate of the chemotactic coupling, we have plotted the fitted $\Omega_\text{C}$ versus the calculated filled micelle gradient, $\partial_r c(t)/c_0$, using \eqnref{eqn:main} (\figref{fig:analysis} (d)). If we assume a linear dependence on the filled micelle gradient~\cite{jin2017_chemotaxis,taktikos2011_modeling}, $\Omega_\text{C}=\kappa\partial_rc/c_0$,  linear regression yields $\kappa\approx\SI{130}{\radian\micro\metre\per\second}$ with $c_0$ on the order of 200 oil molecules per \SI{}{\micro\metre\cubed} (see SI \secref{SIsec:solub}), however, this should be regarded as no more than  an order-of-magnitude estimate due to our limited range of accessible pillar sizes ($\SI{50}{\micro\metre}<R<\SI{250}{\micro\metre})$ and the various approximations in our model. 

In \figref{fig:analysis} (e), we plot the extracted rotational diffusion coefficients. For comparison, we have estimated the average rotational diffusion coefficient of swimmers moving between pillars by computing their mean squared rotation, $D_\text{R}\approx$ \SI{0.05}{\radian\per\second} (See SI~\secref{SIsec:diffusion}). The values for droplets at pillars are quite similar, except for small pillars with chemotactic gradients, which are increased. This could be due to increased rotational noise from the backaction of the inhomogeneous chemotactic field on the surfactant coverage at the interface.

The fact that the rotational diffusion estimates for freely moving and wall attached droplets are comparable is not intuitively obvious, since variations in interfacial tension provide a major source of noise and such variations are affected by the boundary conditions of advective flow and chemical fields. Our data suggests that the differences are small enough to treat the rotational diffusion coeffient of a droplet swimmer as a meaningful physical quantity.

We have shown that from a simple curious phenomenon - droplet swimmers leaving pillars after orbiting them just once - one can independently estimate various quantities like hydrodynamic and chemorepulsive torques, diffusion coefficients and their dependence on wall curvature by a statistical analysis of detention times and their interpretation using an active Brownian swimmer model. This provides valuable insight into droplet swimmers in particular, where many of these quantities are hard to access independently or where their sensitivity to boundary conditions is still open to debate. 
Since the ABP model makes no assumptions about the details of the propulsion process, this assay can also be extended to probe similar quantities in various other microswimmer systems.

We acknowledge Babak V. Hokmabad for experimental, and Fabian Schwarzendahl and Babak Nasouri for theoretical advice. This project was partially funded by the MPI-DS Prandtl internship program and the MaxSynBio initiative.

\bibliography{pillarsPRL.bib}

\begin{thebibliography}{44}%
\makeatletter
\providecommand \@ifxundefined [1]{%
 \@ifx{#1\undefined}
}%
\providecommand \@ifnum [1]{%
 \ifnum #1\expandafter \@firstoftwo
 \else \expandafter \@secondoftwo
 \fi
}%
\providecommand \@ifx [1]{%
 \ifx #1\expandafter \@firstoftwo
 \else \expandafter \@secondoftwo
 \fi
}%
\providecommand \natexlab [1]{#1}%
\providecommand \enquote  [1]{``#1''}%
\providecommand \bibnamefont  [1]{#1}%
\providecommand \bibfnamefont [1]{#1}%
\providecommand \citenamefont [1]{#1}%
\providecommand \href@noop [0]{\@secondoftwo}%
\providecommand \href [0]{\begingroup \@sanitize@url \@href}%
\providecommand \@href[1]{\@@startlink{#1}\@@href}%
\providecommand \@@href[1]{\endgroup#1\@@endlink}%
\providecommand \@sanitize@url [0]{\catcode `\\12\catcode `\$12\catcode
  `\&12\catcode `\#12\catcode `\^12\catcode `\_12\catcode `\%12\relax}%
\providecommand \@@startlink[1]{}%
\providecommand \@@endlink[0]{}%
\providecommand \url  [0]{\begingroup\@sanitize@url \@url }%
\providecommand \@url [1]{\endgroup\@href {#1}{\urlprefix }}%
\providecommand \urlprefix  [0]{URL }%
\providecommand \Eprint [0]{\href }%
\providecommand \doibase [0]{http://dx.doi.org/}%
\providecommand \selectlanguage [0]{\@gobble}%
\providecommand \bibinfo  [0]{\@secondoftwo}%
\providecommand \bibfield  [0]{\@secondoftwo}%
\providecommand \translation [1]{[#1]}%
\providecommand \BibitemOpen [0]{}%
\providecommand \bibitemStop [0]{}%
\providecommand \bibitemNoStop [0]{.\EOS\space}%
\providecommand \EOS [0]{\spacefactor3000\relax}%
\providecommand \BibitemShut  [1]{\csname bibitem#1\endcsname}%
\let\auto@bib@innerbib\@empty
\bibitem [{\citenamefont {Madigan}\ and\ \citenamefont
  {Martinko}(2005)}]{madigan2005_bacterial}%
  \BibitemOpen
  \bibfield  {author} {\bibinfo {author} {\bibfnamefont {M.~T.}\ \bibnamefont
  {Madigan}}\ and\ \bibinfo {author} {\bibfnamefont {J.~M.}\ \bibnamefont
  {Martinko}},\ }in\ \href@noop {} {\emph {\bibinfo {booktitle} {Brock Biology
  of Microorganisms}}}\ (\bibinfo  {publisher} {{Pearson Education, Limited}},\
  \bibinfo {year} {2005})\ Chap.~\bibinfo {chapter} {4}\BibitemShut {NoStop}%
\bibitem [{\citenamefont {Adler}(1966)}]{adler1966_chemotaxis}%
  \BibitemOpen
  \bibfield  {author} {\bibinfo {author} {\bibfnamefont {J.}~\bibnamefont
  {Adler}},\ }\href@noop {} {\bibfield  {journal} {\bibinfo  {journal}
  {Science}\ }\textbf {\bibinfo {volume} {153}},\ \bibinfo {pages} {708}
  (\bibinfo {year} {1966})}\BibitemShut {NoStop}%
\bibitem [{\citenamefont {Hazelbauer}(2012)}]{hazelbauer2012_bacterial}%
  \BibitemOpen
  \bibfield  {author} {\bibinfo {author} {\bibfnamefont {G.~L.}\ \bibnamefont
  {Hazelbauer}},\ }\href@noop {} {\bibfield  {journal} {\bibinfo  {journal}
  {Annu. Rev. Microbiol.}\ }\textbf {\bibinfo {volume} {66}},\ \bibinfo {pages}
  {285} (\bibinfo {year} {2012})}\BibitemShut {NoStop}%
\bibitem [{\citenamefont {Bonner}\ and\ \citenamefont
  {Savage}(1947)}]{bonner1947_evidence}%
  \BibitemOpen
  \bibfield  {author} {\bibinfo {author} {\bibfnamefont {J.~T.}\ \bibnamefont
  {Bonner}}\ and\ \bibinfo {author} {\bibfnamefont {L.~J.}\ \bibnamefont
  {Savage}},\ }\href@noop {} {\bibfield  {journal} {\bibinfo  {journal} {J.
  Exp. Zool.}\ }\textbf {\bibinfo {volume} {106}},\ \bibinfo {pages} {1}
  (\bibinfo {year} {1947})}\BibitemShut {NoStop}%
\bibitem [{\citenamefont {Zhao}\ \emph {et~al.}(2013)\citenamefont {Zhao},
  \citenamefont {Tseng}, \citenamefont {Beckerman}, \citenamefont {Jin},
  \citenamefont {Gibiansky}, \citenamefont {Harrison}, \citenamefont {Luijten},
  \citenamefont {Parsek},\ and\ \citenamefont {Wong}}]{zhao2013_psl}%
  \BibitemOpen
  \bibfield  {author} {\bibinfo {author} {\bibfnamefont {K.}~\bibnamefont
  {Zhao}}, \bibinfo {author} {\bibfnamefont {B.~S.}\ \bibnamefont {Tseng}},
  \bibinfo {author} {\bibfnamefont {B.}~\bibnamefont {Beckerman}}, \bibinfo
  {author} {\bibfnamefont {F.}~\bibnamefont {Jin}}, \bibinfo {author}
  {\bibfnamefont {M.~L.}\ \bibnamefont {Gibiansky}}, \bibinfo {author}
  {\bibfnamefont {J.~J.}\ \bibnamefont {Harrison}}, \bibinfo {author}
  {\bibfnamefont {E.}~\bibnamefont {Luijten}}, \bibinfo {author} {\bibfnamefont
  {M.~R.}\ \bibnamefont {Parsek}}, \ and\ \bibinfo {author} {\bibfnamefont
  {G.~C.~L.}\ \bibnamefont {Wong}},\ }\href@noop {} {\bibfield  {journal}
  {\bibinfo  {journal} {Nature}\ }\textbf {\bibinfo {volume} {497}},\ \bibinfo
  {pages} {388} (\bibinfo {year} {2013})}\BibitemShut {NoStop}%
\bibitem [{\citenamefont {Reid}\ \emph {et~al.}(2012)\citenamefont {Reid},
  \citenamefont {Latty}, \citenamefont {Dussutour},\ and\ \citenamefont
  {Beekman}}]{reid2012_slime}%
  \BibitemOpen
  \bibfield  {author} {\bibinfo {author} {\bibfnamefont {C.~R.}\ \bibnamefont
  {Reid}}, \bibinfo {author} {\bibfnamefont {T.}~\bibnamefont {Latty}},
  \bibinfo {author} {\bibfnamefont {A.}~\bibnamefont {Dussutour}}, \ and\
  \bibinfo {author} {\bibfnamefont {M.}~\bibnamefont {Beekman}},\ }\href@noop
  {} {\bibfield  {journal} {\bibinfo  {journal} {PNAS}\ }\textbf {\bibinfo
  {volume} {109}},\ \bibinfo {pages} {17490} (\bibinfo {year}
  {2012})}\BibitemShut {NoStop}%
\bibitem [{\citenamefont {Spagnolie}\ and\ \citenamefont
  {Lauga}(2012)}]{spagnolie2012_hydrodynamics}%
  \BibitemOpen
  \bibfield  {author} {\bibinfo {author} {\bibfnamefont {S.~E.}\ \bibnamefont
  {Spagnolie}}\ and\ \bibinfo {author} {\bibfnamefont {E.}~\bibnamefont
  {Lauga}},\ }\href@noop {} {\bibfield  {journal} {\bibinfo  {journal} {J.
  Fluid Mech.}\ }\textbf {\bibinfo {volume} {700}},\ \bibinfo {pages} {105}
  (\bibinfo {year} {2012})}\BibitemShut {NoStop}%
\bibitem [{\citenamefont {Berke}\ \emph {et~al.}(2008)\citenamefont {Berke},
  \citenamefont {Turner}, \citenamefont {Berg},\ and\ \citenamefont
  {Lauga}}]{berke2008_hydrodynamic}%
  \BibitemOpen
  \bibfield  {author} {\bibinfo {author} {\bibfnamefont {A.~P.}\ \bibnamefont
  {Berke}}, \bibinfo {author} {\bibfnamefont {L.}~\bibnamefont {Turner}},
  \bibinfo {author} {\bibfnamefont {H.~C.}\ \bibnamefont {Berg}}, \ and\
  \bibinfo {author} {\bibfnamefont {E.}~\bibnamefont {Lauga}},\ }\href@noop {}
  {\bibfield  {journal} {\bibinfo  {journal} {Phys. Rev. Lett.}\ }\textbf
  {\bibinfo {volume} {101}},\ \bibinfo {pages} {038102} (\bibinfo {year}
  {2008})}\BibitemShut {NoStop}%
\bibitem [{\citenamefont {Li}\ and\ \citenamefont
  {Tang}(2009)}]{li2009_accumulation}%
  \BibitemOpen
  \bibfield  {author} {\bibinfo {author} {\bibfnamefont {G.}~\bibnamefont
  {Li}}\ and\ \bibinfo {author} {\bibfnamefont {J.~X.}\ \bibnamefont {Tang}},\
  }\href@noop {} {\bibfield  {journal} {\bibinfo  {journal} {Phys. Rev. Lett.}\
  }\textbf {\bibinfo {volume} {103}},\ \bibinfo {pages} {078101} (\bibinfo
  {year} {2009})}\BibitemShut {NoStop}%
\bibitem [{\citenamefont {{van Teeffelen}}\ and\ \citenamefont
  {L{\"o}wen}(2008)}]{vanteeffelen2008_dynamics}%
  \BibitemOpen
  \bibfield  {author} {\bibinfo {author} {\bibfnamefont {S.}~\bibnamefont {{van
  Teeffelen}}}\ and\ \bibinfo {author} {\bibfnamefont {H.}~\bibnamefont
  {L{\"o}wen}},\ }\href@noop {} {\bibfield  {journal} {\bibinfo  {journal}
  {Phys. Rev. E}\ }\textbf {\bibinfo {volume} {78}},\ \bibinfo {pages} {020101}
  (\bibinfo {year} {2008})}\BibitemShut {NoStop}%
\bibitem [{\citenamefont {Elgeti}\ and\ \citenamefont
  {Gompper}(2013)}]{elgeti2013_wall}%
  \BibitemOpen
  \bibfield  {author} {\bibinfo {author} {\bibfnamefont {J.}~\bibnamefont
  {Elgeti}}\ and\ \bibinfo {author} {\bibfnamefont {G.}~\bibnamefont
  {Gompper}},\ }\href@noop {} {\bibfield  {journal} {\bibinfo  {journal}
  {Europhys. Lett.}\ }\textbf {\bibinfo {volume} {101}},\ \bibinfo {pages}
  {48003} (\bibinfo {year} {2013})}\BibitemShut {NoStop}%
\bibitem [{\citenamefont {Tuval}\ \emph {et~al.}(2005)\citenamefont {Tuval},
  \citenamefont {Cisneros}, \citenamefont {Dombrowski}, \citenamefont
  {Wolgemuth}, \citenamefont {Kessler},\ and\ \citenamefont
  {Goldstein}}]{tuval2005_bacterial}%
  \BibitemOpen
  \bibfield  {author} {\bibinfo {author} {\bibfnamefont {I.}~\bibnamefont
  {Tuval}}, \bibinfo {author} {\bibfnamefont {L.}~\bibnamefont {Cisneros}},
  \bibinfo {author} {\bibfnamefont {C.}~\bibnamefont {Dombrowski}}, \bibinfo
  {author} {\bibfnamefont {C.~W.}\ \bibnamefont {Wolgemuth}}, \bibinfo {author}
  {\bibfnamefont {J.~O.}\ \bibnamefont {Kessler}}, \ and\ \bibinfo {author}
  {\bibfnamefont {R.~E.}\ \bibnamefont {Goldstein}},\ }\href@noop {} {\bibfield
   {journal} {\bibinfo  {journal} {PNAS}\ }\textbf {\bibinfo {volume} {102}},\
  \bibinfo {pages} {2277} (\bibinfo {year} {2005})}\BibitemShut {NoStop}%
\bibitem [{\citenamefont {Jin}\ \emph {et~al.}(2018)\citenamefont {Jin},
  \citenamefont {Vajdi~Hokmabad}, \citenamefont {Baldwin},\ and\ \citenamefont
  {Maass}}]{jin2018_chemotactic}%
  \BibitemOpen
  \bibfield  {author} {\bibinfo {author} {\bibfnamefont {C.}~\bibnamefont
  {Jin}}, \bibinfo {author} {\bibfnamefont {B.}~\bibnamefont {Vajdi~Hokmabad}},
  \bibinfo {author} {\bibfnamefont {K.~A.}\ \bibnamefont {Baldwin}}, \ and\
  \bibinfo {author} {\bibfnamefont {C.~C.}\ \bibnamefont {Maass}},\ }\href@noop
  {} {\bibfield  {journal} {\bibinfo  {journal} {J. Phys.: Condens. Mat.}\
  }\textbf {\bibinfo {volume} {30}},\ \bibinfo {pages} {054003} (\bibinfo
  {year} {2018})}\BibitemShut {NoStop}%
\bibitem [{\citenamefont {Herminghaus}\ \emph {et~al.}(2014)\citenamefont
  {Herminghaus}, \citenamefont {Maass}, \citenamefont {Kr{\"u}ger},
  \citenamefont {Thutupalli}, \citenamefont {Goehring},\ and\ \citenamefont
  {Bahr}}]{herminghaus2014_interfacial}%
  \BibitemOpen
  \bibfield  {author} {\bibinfo {author} {\bibfnamefont {S.}~\bibnamefont
  {Herminghaus}}, \bibinfo {author} {\bibfnamefont {C.~C.}\ \bibnamefont
  {Maass}}, \bibinfo {author} {\bibfnamefont {C.}~\bibnamefont {Kr{\"u}ger}},
  \bibinfo {author} {\bibfnamefont {S.}~\bibnamefont {Thutupalli}}, \bibinfo
  {author} {\bibfnamefont {L.}~\bibnamefont {Goehring}}, \ and\ \bibinfo
  {author} {\bibfnamefont {C.}~\bibnamefont {Bahr}},\ }\href@noop {} {\bibfield
   {journal} {\bibinfo  {journal} {Soft Matter}\ }\textbf {\bibinfo {volume}
  {10}},\ \bibinfo {pages} {7008} (\bibinfo {year} {2014})}\BibitemShut
  {NoStop}%
\bibitem [{\citenamefont {Maass}\ \emph {et~al.}(2016)\citenamefont {Maass},
  \citenamefont {Kr{\"u}ger}, \citenamefont {Herminghaus},\ and\ \citenamefont
  {Bahr}}]{maass2016_swimming}%
  \BibitemOpen
  \bibfield  {author} {\bibinfo {author} {\bibfnamefont {C.~C.}\ \bibnamefont
  {Maass}}, \bibinfo {author} {\bibfnamefont {C.}~\bibnamefont {Kr{\"u}ger}},
  \bibinfo {author} {\bibfnamefont {S.}~\bibnamefont {Herminghaus}}, \ and\
  \bibinfo {author} {\bibfnamefont {C.}~\bibnamefont {Bahr}},\ }\href@noop {}
  {\bibfield  {journal} {\bibinfo  {journal} {Annu Rev Condens Matter Phys}\
  }\textbf {\bibinfo {volume} {7}},\ \bibinfo {pages} {171} (\bibinfo {year}
  {2016})}\BibitemShut {NoStop}%
\bibitem [{\citenamefont {Jin}\ \emph {et~al.}(2017)\citenamefont {Jin},
  \citenamefont {Kr{\"u}ger},\ and\ \citenamefont
  {Maass}}]{jin2017_chemotaxis}%
  \BibitemOpen
  \bibfield  {author} {\bibinfo {author} {\bibfnamefont {C.}~\bibnamefont
  {Jin}}, \bibinfo {author} {\bibfnamefont {C.}~\bibnamefont {Kr{\"u}ger}}, \
  and\ \bibinfo {author} {\bibfnamefont {C.~C.}\ \bibnamefont {Maass}},\
  }\href@noop {} {\bibfield  {journal} {\bibinfo  {journal} {P. Natl. Acad.
  Sci. USA}\ }\textbf {\bibinfo {volume} {114}},\ \bibinfo {pages} {5089}
  (\bibinfo {year} {2017})}\BibitemShut {NoStop}%
\bibitem [{\citenamefont {Candau}\ \emph {et~al.}(1984)\citenamefont {Candau},
  \citenamefont {Hirsch},\ and\ \citenamefont {Zana}}]{candau1984_new}%
  \BibitemOpen
  \bibfield  {author} {\bibinfo {author} {\bibfnamefont {S.}~\bibnamefont
  {Candau}}, \bibinfo {author} {\bibfnamefont {E.}~\bibnamefont {Hirsch}}, \
  and\ \bibinfo {author} {\bibfnamefont {R.}~\bibnamefont {Zana}},\ }\href@noop
  {} {\bibfield  {journal} {\bibinfo  {journal} {J. Phys.}\ }\textbf {\bibinfo
  {volume} {45}},\ \bibinfo {pages} {1263} (\bibinfo {year}
  {1984})}\BibitemShut {NoStop}%
\bibitem [{\citenamefont {Sipos}\ \emph {et~al.}(2015)\citenamefont {Sipos},
  \citenamefont {Nagy}, \citenamefont {Di~Leonardo},\ and\ \citenamefont
  {Galajda}}]{sipos2015_hydrodynamic}%
  \BibitemOpen
  \bibfield  {author} {\bibinfo {author} {\bibfnamefont {O.}~\bibnamefont
  {Sipos}}, \bibinfo {author} {\bibfnamefont {K.}~\bibnamefont {Nagy}},
  \bibinfo {author} {\bibfnamefont {R.}~\bibnamefont {Di~Leonardo}}, \ and\
  \bibinfo {author} {\bibfnamefont {P.}~\bibnamefont {Galajda}},\ }\href@noop
  {} {\bibfield  {journal} {\bibinfo  {journal} {Phys Rev Lett}\ }\textbf
  {\bibinfo {volume} {114}},\ \bibinfo {pages} {258104} (\bibinfo {year}
  {2015})}\BibitemShut {NoStop}%
\bibitem [{\citenamefont {Takagi}\ \emph {et~al.}(2014)\citenamefont {Takagi},
  \citenamefont {Palacci}, \citenamefont {Braunschweig}, \citenamefont
  {Shelley},\ and\ \citenamefont {Zhang}}]{takagi2014_hydrodynamic}%
  \BibitemOpen
  \bibfield  {author} {\bibinfo {author} {\bibfnamefont {D.}~\bibnamefont
  {Takagi}}, \bibinfo {author} {\bibfnamefont {J.}~\bibnamefont {Palacci}},
  \bibinfo {author} {\bibfnamefont {A.~B.}\ \bibnamefont {Braunschweig}},
  \bibinfo {author} {\bibfnamefont {M.~J.}\ \bibnamefont {Shelley}}, \ and\
  \bibinfo {author} {\bibfnamefont {J.}~\bibnamefont {Zhang}},\ }\href@noop {}
  {\bibfield  {journal} {\bibinfo  {journal} {Soft Matter}\ }\textbf {\bibinfo
  {volume} {I}} (\bibinfo {year} {2014})}\BibitemShut {NoStop}%
\bibitem [{\citenamefont {Simmchen}\ \emph {et~al.}(2016)\citenamefont
  {Simmchen}, \citenamefont {Katuri}, \citenamefont {Uspal}, \citenamefont
  {Popescu}, \citenamefont {Tasinkevych},\ and\ \citenamefont
  {S{\'a}nchez}}]{simmchen2016_topographical}%
  \BibitemOpen
  \bibfield  {author} {\bibinfo {author} {\bibfnamefont {J.}~\bibnamefont
  {Simmchen}}, \bibinfo {author} {\bibfnamefont {J.}~\bibnamefont {Katuri}},
  \bibinfo {author} {\bibfnamefont {W.~E.}\ \bibnamefont {Uspal}}, \bibinfo
  {author} {\bibfnamefont {M.~N.}\ \bibnamefont {Popescu}}, \bibinfo {author}
  {\bibfnamefont {M.}~\bibnamefont {Tasinkevych}}, \ and\ \bibinfo {author}
  {\bibfnamefont {S.}~\bibnamefont {S{\'a}nchez}},\ }\href@noop {} {\bibfield
  {journal} {\bibinfo  {journal} {Nat. Commun.}\ }\textbf {\bibinfo {volume}
  {7}},\ \bibinfo {pages} {10598} (\bibinfo {year} {2016})}\BibitemShut
  {NoStop}%
\bibitem [{\citenamefont {Schaar}\ \emph {et~al.}(2015)\citenamefont {Schaar},
  \citenamefont {Z{\"o}ttl},\ and\ \citenamefont
  {Stark}}]{schaar2015_detention}%
  \BibitemOpen
  \bibfield  {author} {\bibinfo {author} {\bibfnamefont {K.}~\bibnamefont
  {Schaar}}, \bibinfo {author} {\bibfnamefont {A.}~\bibnamefont {Z{\"o}ttl}}, \
  and\ \bibinfo {author} {\bibfnamefont {H.}~\bibnamefont {Stark}},\
  }\href@noop {} {\bibfield  {journal} {\bibinfo  {journal} {Phys. Rev. Lett.}\
  }\textbf {\bibinfo {volume} {115}},\ \bibinfo {pages} {038101} (\bibinfo
  {year} {2015})}\BibitemShut {NoStop}%
\bibitem [{\citenamefont {Spagnolie}\ \emph {et~al.}(2015)\citenamefont
  {Spagnolie}, \citenamefont {{Moreno-Flores}}, \citenamefont {Bartolo},\ and\
  \citenamefont {Lauga}}]{spagnolie2015_geometric}%
  \BibitemOpen
  \bibfield  {author} {\bibinfo {author} {\bibfnamefont {S.~E.}\ \bibnamefont
  {Spagnolie}}, \bibinfo {author} {\bibfnamefont {G.~R.}\ \bibnamefont
  {{Moreno-Flores}}}, \bibinfo {author} {\bibfnamefont {D.}~\bibnamefont
  {Bartolo}}, \ and\ \bibinfo {author} {\bibfnamefont {E.}~\bibnamefont
  {Lauga}},\ }\href@noop {} {\bibfield  {journal} {\bibinfo  {journal} {Soft
  Matter}\ }\textbf {\bibinfo {volume} {11}},\ \bibinfo {pages} {3396}
  (\bibinfo {year} {2015})}\BibitemShut {NoStop}%
\bibitem [{\citenamefont {Romanczuk}\ \emph {et~al.}(2012)\citenamefont
  {Romanczuk}, \citenamefont {B{\"a}r}, \citenamefont {Ebeling}, \citenamefont
  {Lindner},\ and\ \citenamefont {{Schimansky-Geier}}}]{romanczuk2012_active}%
  \BibitemOpen
  \bibfield  {author} {\bibinfo {author} {\bibfnamefont {P.}~\bibnamefont
  {Romanczuk}}, \bibinfo {author} {\bibfnamefont {M.}~\bibnamefont {B{\"a}r}},
  \bibinfo {author} {\bibfnamefont {W.}~\bibnamefont {Ebeling}}, \bibinfo
  {author} {\bibfnamefont {B.}~\bibnamefont {Lindner}}, \ and\ \bibinfo
  {author} {\bibfnamefont {L.}~\bibnamefont {{Schimansky-Geier}}},\ }\href@noop
  {} {\bibfield  {journal} {\bibinfo  {journal} {Eur. Phys. J. Spec. Top.}\
  }\textbf {\bibinfo {volume} {202}},\ \bibinfo {pages} {1} (\bibinfo {year}
  {2012})}\BibitemShut {NoStop}%
\bibitem [{\citenamefont {Schmitt}\ and\ \citenamefont
  {Stark}(2016)}]{schmitt2016_active}%
  \BibitemOpen
  \bibfield  {author} {\bibinfo {author} {\bibfnamefont {M.}~\bibnamefont
  {Schmitt}}\ and\ \bibinfo {author} {\bibfnamefont {H.}~\bibnamefont
  {Stark}},\ }\href@noop {} {\bibfield  {journal} {\bibinfo  {journal} {Eur.
  Phys. J. E}\ }\textbf {\bibinfo {volume} {39}},\ \bibinfo {pages} {80}
  (\bibinfo {year} {2016})}\BibitemShut {NoStop}%
\bibitem [{\citenamefont {Z{\"o}ttl}\ and\ \citenamefont
  {Stark}(2016)}]{zottl2016_emergent}%
  \BibitemOpen
  \bibfield  {author} {\bibinfo {author} {\bibfnamefont {A.}~\bibnamefont
  {Z{\"o}ttl}}\ and\ \bibinfo {author} {\bibfnamefont {H.}~\bibnamefont
  {Stark}},\ }\href@noop {} {\bibfield  {journal} {\bibinfo  {journal} {J.
  Phys.: Condens. Matter}\ }\textbf {\bibinfo {volume} {28}},\ \bibinfo {pages}
  {253001} (\bibinfo {year} {2016})}\BibitemShut {NoStop}%
\bibitem [{\citenamefont {Kr{\"u}ger}\ \emph {et~al.}(2016)\citenamefont
  {Kr{\"u}ger}, \citenamefont {Kl{\"o}s}, \citenamefont {Bahr},\ and\
  \citenamefont {Maass}}]{kruger2016_curling}%
  \BibitemOpen
  \bibfield  {author} {\bibinfo {author} {\bibfnamefont {C.}~\bibnamefont
  {Kr{\"u}ger}}, \bibinfo {author} {\bibfnamefont {G.}~\bibnamefont
  {Kl{\"o}s}}, \bibinfo {author} {\bibfnamefont {C.}~\bibnamefont {Bahr}}, \
  and\ \bibinfo {author} {\bibfnamefont {C.~C.}\ \bibnamefont {Maass}},\
  }\href@noop {} {\bibfield  {journal} {\bibinfo  {journal} {Phys. Rev. Lett.}\
  }\textbf {\bibinfo {volume} {117}},\ \bibinfo {pages} {048003} (\bibinfo
  {year} {2016})}\BibitemShut {NoStop}%
\bibitem [{\citenamefont {Grundmann}(2009)}]{grundmann2009_boundary}%
  \BibitemOpen
  \bibfield  {author} {\bibinfo {author} {\bibfnamefont {R.}~\bibnamefont
  {Grundmann}},\ }in\ \href@noop {} {\emph {\bibinfo {booktitle} {Computational
  {{Fluid Dynamics}}}}},\ \bibinfo {editor} {edited by\ \bibinfo {editor}
  {\bibfnamefont {J.~F.}\ \bibnamefont {Wendt}}}\ (\bibinfo  {publisher}
  {{Springer Berlin Heidelberg}},\ \bibinfo {address} {{Berlin, Heidelberg}},\
  \bibinfo {year} {2009})\ pp.\ \bibinfo {pages} {153--181}\BibitemShut
  {NoStop}%
\bibitem [{\citenamefont {Taktikos}\ \emph {et~al.}(2011)\citenamefont
  {Taktikos}, \citenamefont {Zaburdaev},\ and\ \citenamefont
  {Stark}}]{taktikos2011_modeling}%
  \BibitemOpen
  \bibfield  {author} {\bibinfo {author} {\bibfnamefont {J.}~\bibnamefont
  {Taktikos}}, \bibinfo {author} {\bibfnamefont {V.}~\bibnamefont {Zaburdaev}},
  \ and\ \bibinfo {author} {\bibfnamefont {H.}~\bibnamefont {Stark}},\
  }\href@noop {} {\bibfield  {journal} {\bibinfo  {journal} {Phys Rev E}\
  }\textbf {\bibinfo {volume} {84}},\ \bibinfo {pages} {041924} (\bibinfo
  {year} {2011})}\BibitemShut {NoStop}%
\bibitem [{\citenamefont {Qin}\ \emph {et~al.}(2010)\citenamefont {Qin},
  \citenamefont {Xia},\ and\ \citenamefont {Whitesides}}]{qin2010_soft}%
  \BibitemOpen
  \bibfield  {author} {\bibinfo {author} {\bibfnamefont {D.}~\bibnamefont
  {Qin}}, \bibinfo {author} {\bibfnamefont {Y.}~\bibnamefont {Xia}}, \ and\
  \bibinfo {author} {\bibfnamefont {G.~M.}\ \bibnamefont {Whitesides}},\
  }\href@noop {} {\bibfield  {journal} {\bibinfo  {journal} {Nat. Protoc.}\
  }\textbf {\bibinfo {volume} {5}},\ \bibinfo {pages} {491} (\bibinfo {year}
  {2010})}\BibitemShut {NoStop}%
\bibitem [{\citenamefont {Thorsen}\ \emph {et~al.}(2001)\citenamefont
  {Thorsen}, \citenamefont {Roberts}, \citenamefont {Arnold},\ and\
  \citenamefont {Quake}}]{thorsen2001_dynamic}%
  \BibitemOpen
  \bibfield  {author} {\bibinfo {author} {\bibfnamefont {T.}~\bibnamefont
  {Thorsen}}, \bibinfo {author} {\bibfnamefont {R.~W.}\ \bibnamefont
  {Roberts}}, \bibinfo {author} {\bibfnamefont {F.~H.}\ \bibnamefont {Arnold}},
  \ and\ \bibinfo {author} {\bibfnamefont {S.~R.}\ \bibnamefont {Quake}},\
  }\href@noop {} {\bibfield  {journal} {\bibinfo  {journal} {Phys. Rev. Lett.}\
  }\textbf {\bibinfo {volume} {86}},\ \bibinfo {pages} {4163} (\bibinfo {year}
  {2001})}\BibitemShut {NoStop}%
\bibitem [{\citenamefont {Crocker}\ and\ \citenamefont
  {Grier}(1996)}]{crocker1996_methods}%
  \BibitemOpen
  \bibfield  {author} {\bibinfo {author} {\bibfnamefont {J.~C.}\ \bibnamefont
  {Crocker}}\ and\ \bibinfo {author} {\bibfnamefont {D.~G.}\ \bibnamefont
  {Grier}},\ }\href@noop {} {\bibfield  {journal} {\bibinfo  {journal} {J
  Colloid Interface Sci}\ }\textbf {\bibinfo {volume} {179}},\ \bibinfo {pages}
  {298} (\bibinfo {year} {1996})}\BibitemShut {NoStop}%
\bibitem [{\citenamefont {Kr{\"u}ger}(2016)}]{kruger2016_liquid}%
  \BibitemOpen
  \bibfield  {author} {\bibinfo {author} {\bibfnamefont {C.}~\bibnamefont
  {Kr{\"u}ger}},\ }\emph {\bibinfo {title} {Liquid {{Crystal Microswimmers}} -
  from Single Entities to Collective Dynamics}},\ \href@noop {} {Ph.D.
  thesis},\ \bibinfo  {school} {University of G{\"o}ttingen} (\bibinfo {year}
  {2016})\BibitemShut {NoStop}%
\bibitem [{\citenamefont {Howse}\ \emph {et~al.}(2007)\citenamefont {Howse},
  \citenamefont {Jones}, \citenamefont {Ryan}, \citenamefont {Gough},
  \citenamefont {Vafabakhsh},\ and\ \citenamefont
  {Golestanian}}]{howse2007_self-motile}%
  \BibitemOpen
  \bibfield  {author} {\bibinfo {author} {\bibfnamefont {J.~R.}\ \bibnamefont
  {Howse}}, \bibinfo {author} {\bibfnamefont {R.~A.}\ \bibnamefont {Jones}},
  \bibinfo {author} {\bibfnamefont {A.~J.}\ \bibnamefont {Ryan}}, \bibinfo
  {author} {\bibfnamefont {T.}~\bibnamefont {Gough}}, \bibinfo {author}
  {\bibfnamefont {R.}~\bibnamefont {Vafabakhsh}}, \ and\ \bibinfo {author}
  {\bibfnamefont {R.}~\bibnamefont {Golestanian}},\ }\href@noop {} {\bibfield
  {journal} {\bibinfo  {journal} {Phys. Rev. Lett.}\ }\textbf {\bibinfo
  {volume} {99}},\ \bibinfo {pages} {048102} (\bibinfo {year}
  {2007})}\BibitemShut {NoStop}%
\bibitem [{\citenamefont {Saragosti}\ \emph {et~al.}(2012)\citenamefont
  {Saragosti}, \citenamefont {Silberzan},\ and\ \citenamefont
  {Buguin}}]{saragosti2012_modeling}%
  \BibitemOpen
  \bibfield  {author} {\bibinfo {author} {\bibfnamefont {J.}~\bibnamefont
  {Saragosti}}, \bibinfo {author} {\bibfnamefont {P.}~\bibnamefont
  {Silberzan}}, \ and\ \bibinfo {author} {\bibfnamefont {A.}~\bibnamefont
  {Buguin}},\ }\href@noop {} {\bibfield  {journal} {\bibinfo  {journal} {PLOS
  ONE}\ }\textbf {\bibinfo {volume} {7}},\ \bibinfo {pages} {e35412} (\bibinfo
  {year} {2012})}\BibitemShut {NoStop}%
\bibitem [{\citenamefont {Drescher}\ \emph {et~al.}(2011)\citenamefont
  {Drescher}, \citenamefont {Dunkel}, \citenamefont {Cisneros}, \citenamefont
  {Ganguly},\ and\ \citenamefont {Goldstein}}]{drescher2011_fluid}%
  \BibitemOpen
  \bibfield  {author} {\bibinfo {author} {\bibfnamefont {K.}~\bibnamefont
  {Drescher}}, \bibinfo {author} {\bibfnamefont {J.}~\bibnamefont {Dunkel}},
  \bibinfo {author} {\bibfnamefont {L.~H.}\ \bibnamefont {Cisneros}}, \bibinfo
  {author} {\bibfnamefont {S.}~\bibnamefont {Ganguly}}, \ and\ \bibinfo
  {author} {\bibfnamefont {R.~E.}\ \bibnamefont {Goldstein}},\ }\href@noop {}
  {\bibfield  {journal} {\bibinfo  {journal} {PNAS}\ }\textbf {\bibinfo
  {volume} {108}},\ \bibinfo {pages} {10940} (\bibinfo {year}
  {2011})}\BibitemShut {NoStop}%
\bibitem [{\citenamefont {Konotop}\ and\ \citenamefont
  {Vazquez}(1994)}]{konotop1994_nonlinear}%
  \BibitemOpen
  \bibfield  {author} {\bibinfo {author} {\bibfnamefont {V.~V.}\ \bibnamefont
  {Konotop}}\ and\ \bibinfo {author} {\bibfnamefont {L.}~\bibnamefont
  {Vazquez}},\ }\href@noop {} {\emph {\bibinfo {title} {Nonlinear {{Random
  Waves}}}}}\ (\bibinfo  {publisher} {{World Scientific}},\ \bibinfo {year}
  {1994})\BibitemShut {NoStop}%
\bibitem [{\citenamefont {Sevilla}\ and\ \citenamefont
  {Sandoval}(2015)}]{sevilla2015_smoluchowski}%
  \BibitemOpen
  \bibfield  {author} {\bibinfo {author} {\bibfnamefont {F.~J.}\ \bibnamefont
  {Sevilla}}\ and\ \bibinfo {author} {\bibfnamefont {M.}~\bibnamefont
  {Sandoval}},\ }\href@noop {} {\bibfield  {journal} {\bibinfo  {journal}
  {Phys. Rev. E}\ }\textbf {\bibinfo {volume} {91}},\ \bibinfo {pages} {052150}
  (\bibinfo {year} {2015})}\BibitemShut {NoStop}%
\bibitem [{\citenamefont {Sevilla}\ and\ \citenamefont
  {G{\'o}mez~Nava}(2014)}]{sevilla2014_theory}%
  \BibitemOpen
  \bibfield  {author} {\bibinfo {author} {\bibfnamefont {F.~J.}\ \bibnamefont
  {Sevilla}}\ and\ \bibinfo {author} {\bibfnamefont {L.~A.}\ \bibnamefont
  {G{\'o}mez~Nava}},\ }\href@noop {} {\bibfield  {journal} {\bibinfo  {journal}
  {Phys. Rev. E}\ }\textbf {\bibinfo {volume} {90}},\ \bibinfo {pages} {022130}
  (\bibinfo {year} {2014})}\BibitemShut {NoStop}%
\bibitem [{\citenamefont {Kampen}(1992)}]{kampen1992_stochastic}%
  \BibitemOpen
  \bibfield  {author} {\bibinfo {author} {\bibfnamefont {N.~G.~V.}\
  \bibnamefont {Kampen}},\ }\href@noop {} {\emph {\bibinfo {title} {Stochastic
  {{Processes}} in {{Physics}} and {{Chemistry}}}}}\ (\bibinfo  {publisher}
  {{Elsevier}},\ \bibinfo {year} {1992})\BibitemShut {NoStop}%
\bibitem [{\citenamefont {Klyatskin}(2014)}]{klyatskin2014_stochastic}%
  \BibitemOpen
  \bibfield  {author} {\bibinfo {author} {\bibfnamefont {V.~I.}\ \bibnamefont
  {Klyatskin}},\ }\href@noop {} {\emph {\bibinfo {title} {Stochastic
  {{Equations}}: {{Theory}} and {{Applications}} in {{Acoustics}},
  {{Hydrodynamics}}, {{Magnetohydrodynamics}}, and {{Radiophysics}}, {{Volume}}
  2: {{Coherent Phenomena}} in {{Stochastic Dynamic Systems}}}}}\ (\bibinfo
  {publisher} {{Springer}},\ \bibinfo {year} {2014})\BibitemShut {NoStop}%
\bibitem [{\citenamefont {Frank}(2005)}]{frank2005_delay}%
  \BibitemOpen
  \bibfield  {author} {\bibinfo {author} {\bibfnamefont {T.~D.}\ \bibnamefont
  {Frank}},\ }\href@noop {} {\bibfield  {journal} {\bibinfo  {journal} {Phys.
  Rev. E}\ }\textbf {\bibinfo {volume} {71}},\ \bibinfo {pages} {031106}
  (\bibinfo {year} {2005})}\BibitemShut {NoStop}%
\bibitem [{\citenamefont {Novikov}(1965)}]{novikov1965_functionals}%
  \BibitemOpen
  \bibfield  {author} {\bibinfo {author} {\bibfnamefont {E.~A.}\ \bibnamefont
  {Novikov}},\ }\href@noop {} {\bibfield  {journal} {\bibinfo  {journal} {Sov.
  Phys. JETP}\ }\textbf {\bibinfo {volume} {20}},\ \bibinfo {pages} {1290}
  (\bibinfo {year} {1965})}\BibitemShut {NoStop}%
\bibitem [{\citenamefont {{Zinn-Justin}}(2002)}]{zinn-justin2002_quantum}%
  \BibitemOpen
  \bibfield  {author} {\bibinfo {author} {\bibfnamefont {J.}~\bibnamefont
  {{Zinn-Justin}}},\ }\href@noop {} {\emph {\bibinfo {title} {Quantum {{Field
  Theory}} and {{Critical Phenomena}}}}}\ (\bibinfo  {publisher} {{Oxford
  University Press}},\ \bibinfo {year} {2002})\BibitemShut {NoStop}%
\bibitem [{\citenamefont {Vachier}\ and\ \citenamefont
  {Mazza}(2019)}]{vachier2019_dynamics}%
  \BibitemOpen
  \bibfield  {author} {\bibinfo {author} {\bibfnamefont {J.}~\bibnamefont
  {Vachier}}\ and\ \bibinfo {author} {\bibfnamefont {M.~G.}\ \bibnamefont
  {Mazza}},\ }\href@noop {} {\bibfield  {journal} {\bibinfo  {journal} {Eur.
  Phys. J. E}\ }\textbf {\bibinfo {volume} {42}},\ \bibinfo {pages} {11}
  (\bibinfo {year} {2019})}\BibitemShut {NoStop}%
\end{thebibliography}%

\onecolumngrid

\clearpage
\beginsupplement
{\centering\Large{\textbf{Supporting Information}}}
\section{Experimental system}\label{SIsec:methods}
The microfluidic devices are fabricated with standard soft lithography techniques\cite{qin2010_soft}. 
We design photomasks in a 2D AutoCad application and have them printed as a high-resolution emulsion film by an external company (128,000 dots per inch; JD Photo-Tools). 
From the printed photomask, we synthesize a SU-8 (Micro Resist Technology) photoresist mold on a Si wafer ((Wafer World Inc.) in a clean room environment using UV lithography.
The mold is then used to generate PDMS (polydimethylsiloxane; Sylgard 184; Dow Corning) imprints. After punching in fluid inlets and outlets, the imprints are bonded to glass slides.
Covalent bonding between PDMS and glass is achieved by pretreating all surfaces in an air plasma (Pico P100-8; Diener Electronic GmbH + Co. KG) for 30 s.
We produce monodisperse droplets in flow-focusing microfluidic devices~\cite{thorsen2001_dynamic} at high production rates (\SIrange{10}{30}{\per\second}). Droplet radii $a$ can be tuned by flow rates and channel geometries, we used $a=\SI{50}{\micro\metre}$.
The syringes (Braun) are mounted on a precision microfluidic pump (NEM-B101-02B; Cetoni GmbH) and connected to the inlets and outlets with Teflon tubing (39241; Novodirect GmbH). 
To generate oil-in-water emulsions, the primarily hydrophobic PDMS surfaces have to be hydrophilized, first by activation via a  1:1 volumetric mixture of \ce{H2O2} /HCl, then filling the channels with a silanization solution \ce{(C2H2O)_nC7H18O4Si} for 30 min, and finally rinsing them with milli-Q water.

\noindent\begin{minipage}{\textwidth}\begin{center}
    \includegraphics[width=.47\textwidth]{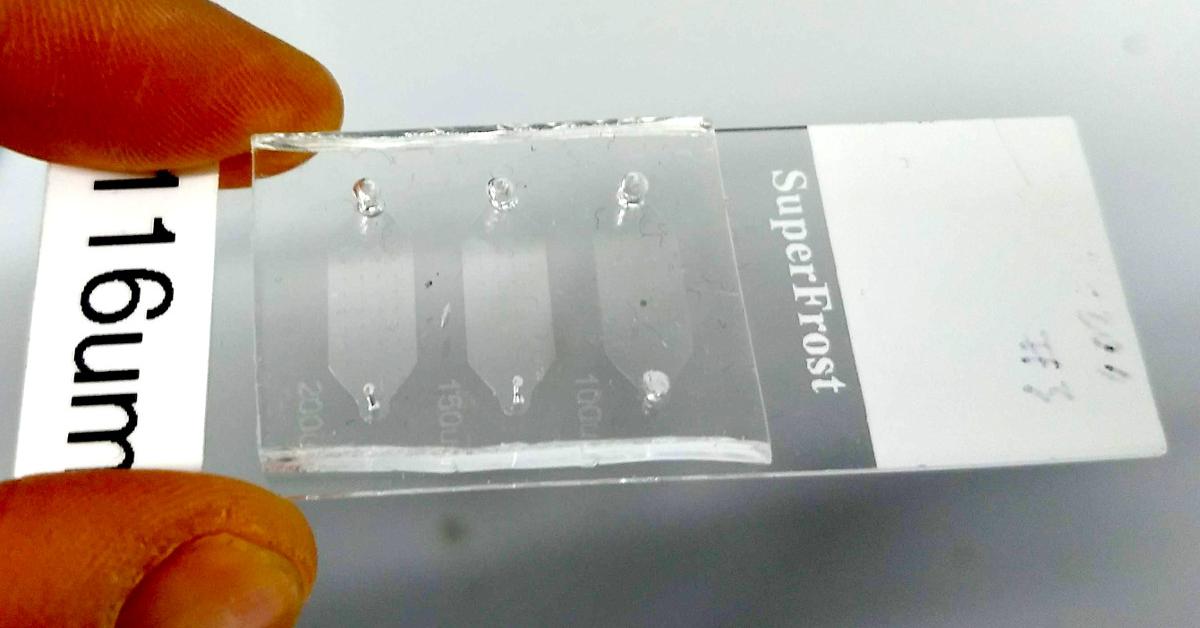}\hspace{5mm}\includegraphics[width=.38\textwidth]{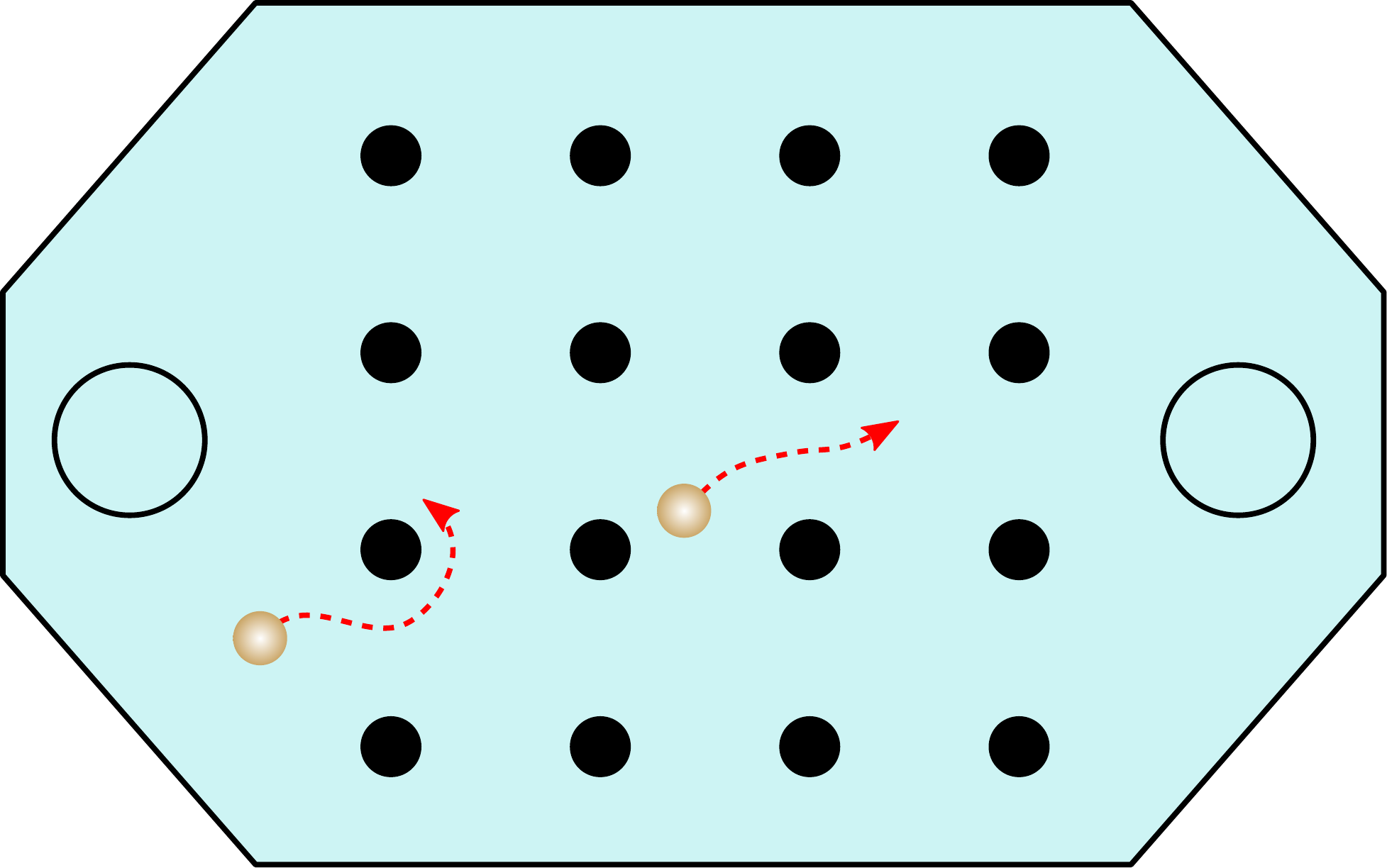}
    \captionof{figure}{left: microfluidic pillar array used in the experiment, cell height \SI{116}{\micro\metre}. right: pillar array schematic}
    \label{SIfig:cell}
\end{center}\end{minipage}

For droplet production and storage we use a submicellar $0.1\,\text{~wt.\%}$ aqueous solution of the ionic surfactant tetradecyltrimethylammonium bromide (TTAB, critical micellar concentration  $c_\mathrm{cmc}=0.13\,\text{~wt.\%}$). The concentration is sufficient to inhibit coalescence, but not high enough for oil solubilisation or droplet swimming.
The oil phase is a mixture of the nematogen 4-pentyl-4'-cyano-biphenyl (5CB)  and 1-Bromopentadecane (BPD) with a volume ratio of $10:1$. Pure 5CB is nematic at room temperature and deformation of the director field inside the droplet triggers helical swimming; the BPD mixture is isotropic and produces persistent Brownian swimmers.

The TTAB concentration in the experimental media is $7.5\,\text{wt} \%$, approx. $60\cdot c_\mathrm{cmc}$.
At the beginning of each experiment,  we add $\approx{\SI{0.3}{\ul}}$ of stock oil emulsion to a \SI{10}{\ul}  $7.5\,\text{wt} \%$ TTAB solution, and pipette the mixture (swimmer density ($n<\SI{5}{/\ul}$)) into the experimental container. 
All experiments are carried out in quasi 2D microfluidic containers. Typical PDMS cells used for experimental observations and a respective schematic drawing are shown in \figref{SIfig:cell}.

We observe the droplets on an Olympus IX-81 bright field microscope under 2x magnification. Images and movies are recorded by a commercial digital camera (Canon EOS 600D) at four frames per second.
For one pillar size, we typically have data from more than $20$ movies recorded over \SIrange{5}{10}{\minute} with $5$ to $20$ swimmers per cell and $1$ to $200$ individual pillar interactions per run.
In the main manuscript (MS), we only included interactions at ``clean'' pillars, i.e.\ the first interaction with each pillar for each run, to avoid possible chemical contamination.  Additional data for all recorded interactions are included in this supplement. 
Sample sizes used for the detention time statistics are listed in Table~\ref{SItbl:sample}.

\noindent\begin{minipage}{\textwidth}\begin{center} 
  \begin{tabular*}{0.75\textwidth}{@{\extracolsep{\fill}} l | c c c c c c c}
    \hline
    $R$ (\si{\um}) & 30 & 50 & 75 & 100 & 250 & 400 & 500\\
    \hline
    Total Sample Size & 319 & 2373 & 1565 & 1176 & 1228 & 789 & 119 \\
    \hline
    First-hit Sample Size & 97 & 821 & 504 & 500 & 354 & 60 & 30\\
    \hline
  \end{tabular*}
   \captionof{table}{\ Sample sizes: number of recorded interactions per pillar size}
  \label{SItbl:sample}
\end{center}\end{minipage}
\section{Numerical data analysis}
We process video microscopy data, tracking the swimmers and extracting trajectories, using software written in-house in Python/openCV, based on a Crocker-Grier type algorithm~\cite{crocker1996_methods}.
We then identify all trajectory segments corresponding to pillar attachment (by Python script) and bin them by detention time $t_\text{d}$.
\subsection{Criteria for trajectory segmentation}\label{SIsec:criteria}

As illustrated in \figref{fig:trail} in the MS, when a swimmer attaches or detaches from a pillar, it reorients and $\theta$ decreases or increases suddenly, leading to a peak in $|\partial_t \theta|$. We use this fact to numerically identify the trajectory segments corresponding to pillar attachment with the following procedure:
On all recorded trajectories, we first apply distance threshold with respect to the individual pillar centres, identifying segments close to pillars. We then calculate $\alpha$ and $\theta$ along each segment (\figref{SIfig:coord}). The attaching and detaching points are decided by combining the thresholds $\tan(\theta)<1$ for $\theta$ and $| \tan \theta (t_0+0.25 s) - \tan \theta (t_0)| < 0.5$ for $|\partial_t \theta|$, which results in an escape angle of \SI{0.962}{\radian}. We have tested the robustness of this procedure by plotting and evaluating segments for multiple sample trajectories (cf.~\figref{SIfig:traj-analysis}).
\noindent\begin{minipage}{\textwidth}\begin{center}
  \includegraphics[width=.9\textwidth]{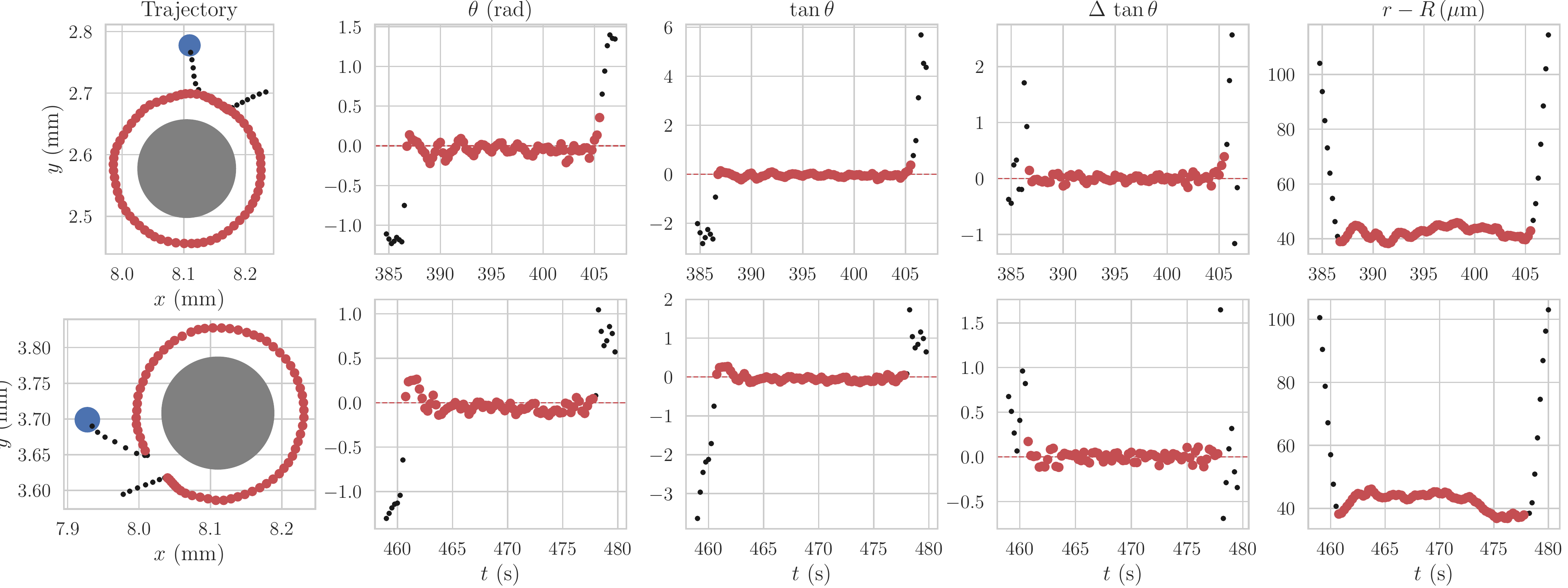}
  \captionof{figure}{Segmentation criteria: one non-crossing and one self-crossing interaction with numerically identified attachment segments marked in red.}
  \label{SIfig:traj-analysis}
\end{center}\end{minipage}

\subsection{Statistical quantities}
We transform the recorded Cartesian $(x,y)$ coordinates to the polar $(r,\alpha)$ system defined in \figref{fig:scheme} in the MS. Individual orbits can be identified by a path integral of $\alpha$ along the trajectory segment, detention times are given by the fixed frame rate of $\SI{4}{\hertz}$.
Detention lengths $s$ are calculated by integrating the distance between frames.
\subsection{Data sets and histograms}\label{SIsec:histograms}
This section contains raw histograms for a larger range of pillar sizes, $R\in\{30,50,75,100,250,\allowbreak 400,500\}\SI{}{\micro\metre}$ 
 and explains our choice of data representation in \figref{fig:hist-fit}, MS.
 
We note that the data sets for $R<\SI{50}{\um}$ and $R>\SI{250}{\um}$ represent pure scattering and trapping regimes and are therefore not usable in our analysis of chemotactically biased detachment.

We can choose between binning by detention length in the number of orbits (\figref{SIfig:hist-all-circ}), and the detention time in seconds (\figref{SIfig:hist-all-time}). Binning by orbit or length clearly marks the onset of chemotactic repulsion, while binning by time is preferable if we want to fit to the time correlated stochastic noise in our Langevin model. 
The two approaches are only equivalent if the droplet speed is the same and absolutely constant for all swimmers; in practice, the binning by time will result in a broadening of peaks by the speed distribution, as shown in~\figref{SIfig:hist-all-time}. To correct for this error, we have chosen to normalise all second-orbit segments to the point in time when each trajectory crosses itself. We have also excluded all interactions with pillars that have previously been hit by swimmers and might have residual chemorepellent contamination (\figref{SIfig:hist-first-time}). 

\noindent\begin{minipage}{\textwidth}\begin{center}
  \includegraphics[width=.9\textwidth]{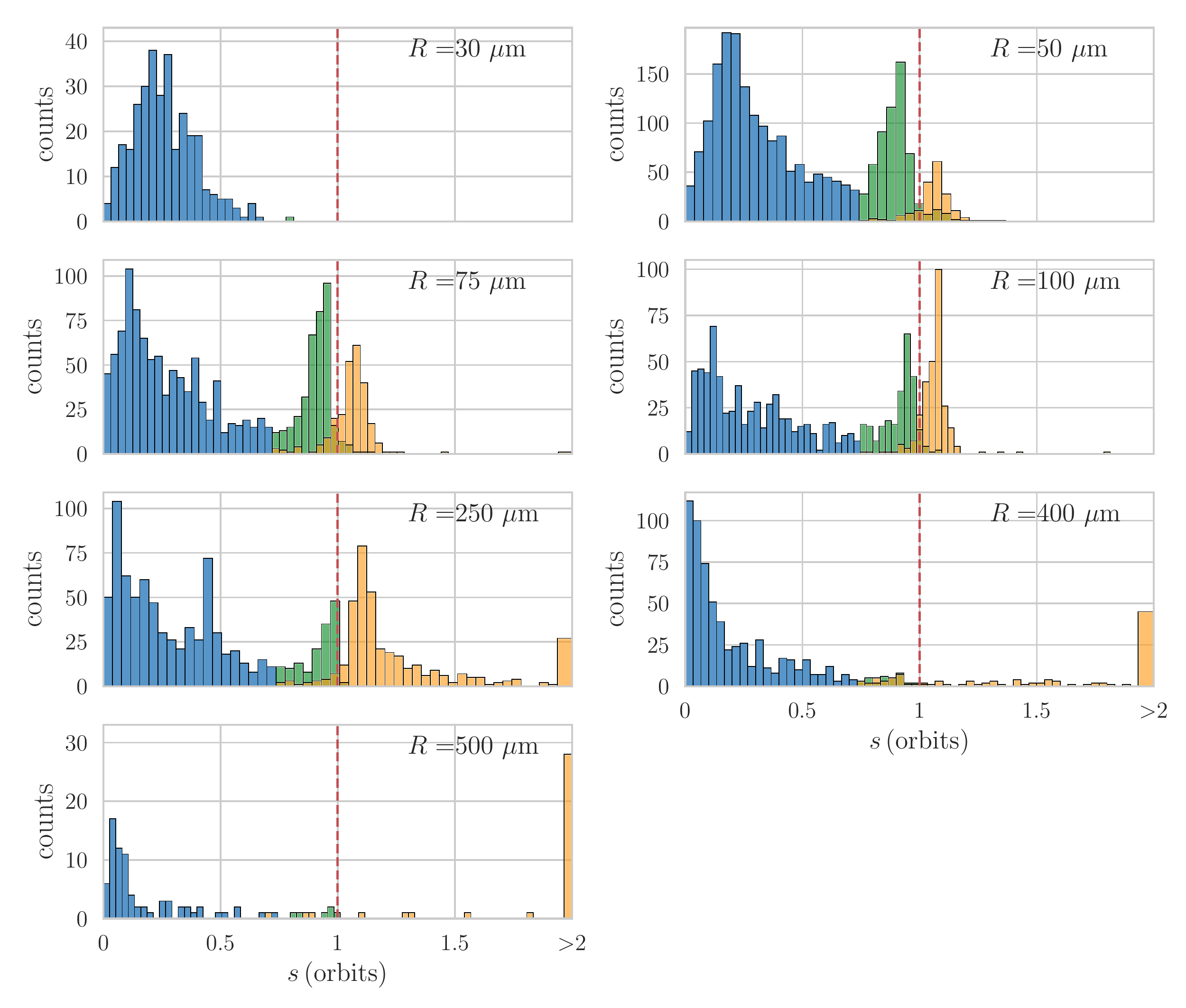}
  \captionof{figure}{Histograms of the detention length measured by orbits for varying pillar radius. The last column bins all lengths in excess of two orbits. The data set includes all recorded interactions. Blue bars mark scattering, 
  green non-crossing and yellow self-crossing interactions (see also \figref{fig:sample-hist}, MS)}.
  \label{SIfig:hist-all-circ}
\end{center}\end{minipage}

\noindent\begin{minipage}{\textwidth}\begin{center}
  \includegraphics[width=.9\textwidth]{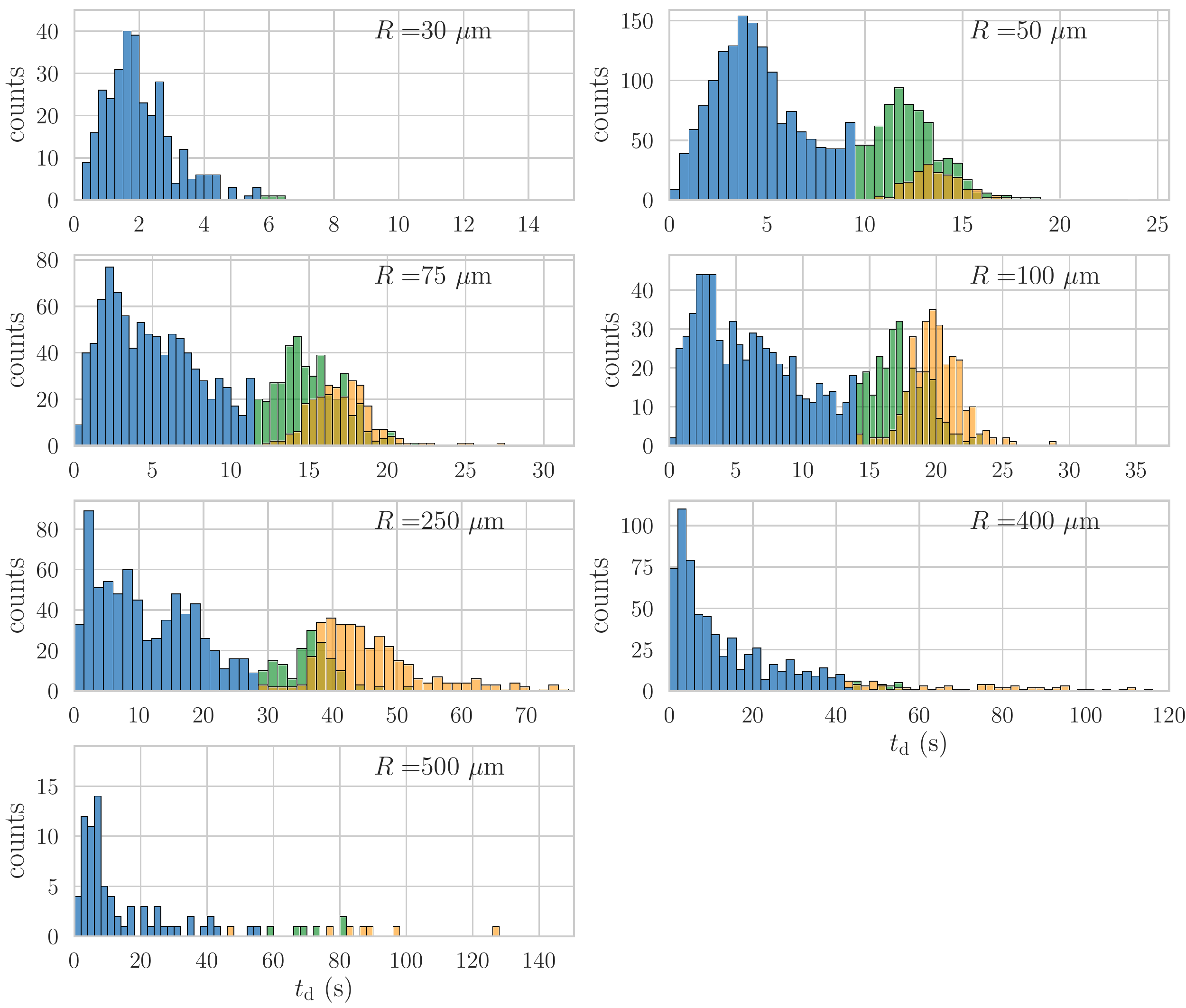}
  \captionof{figure}{Histograms of the detention times at pillars of varying radius. The data set includes all recorded interactions.}
  \label{SIfig:hist-all-time}
\end{center}\end{minipage}

\noindent\begin{minipage}{\textwidth}\begin{center}
  \includegraphics[width=.9\textwidth]{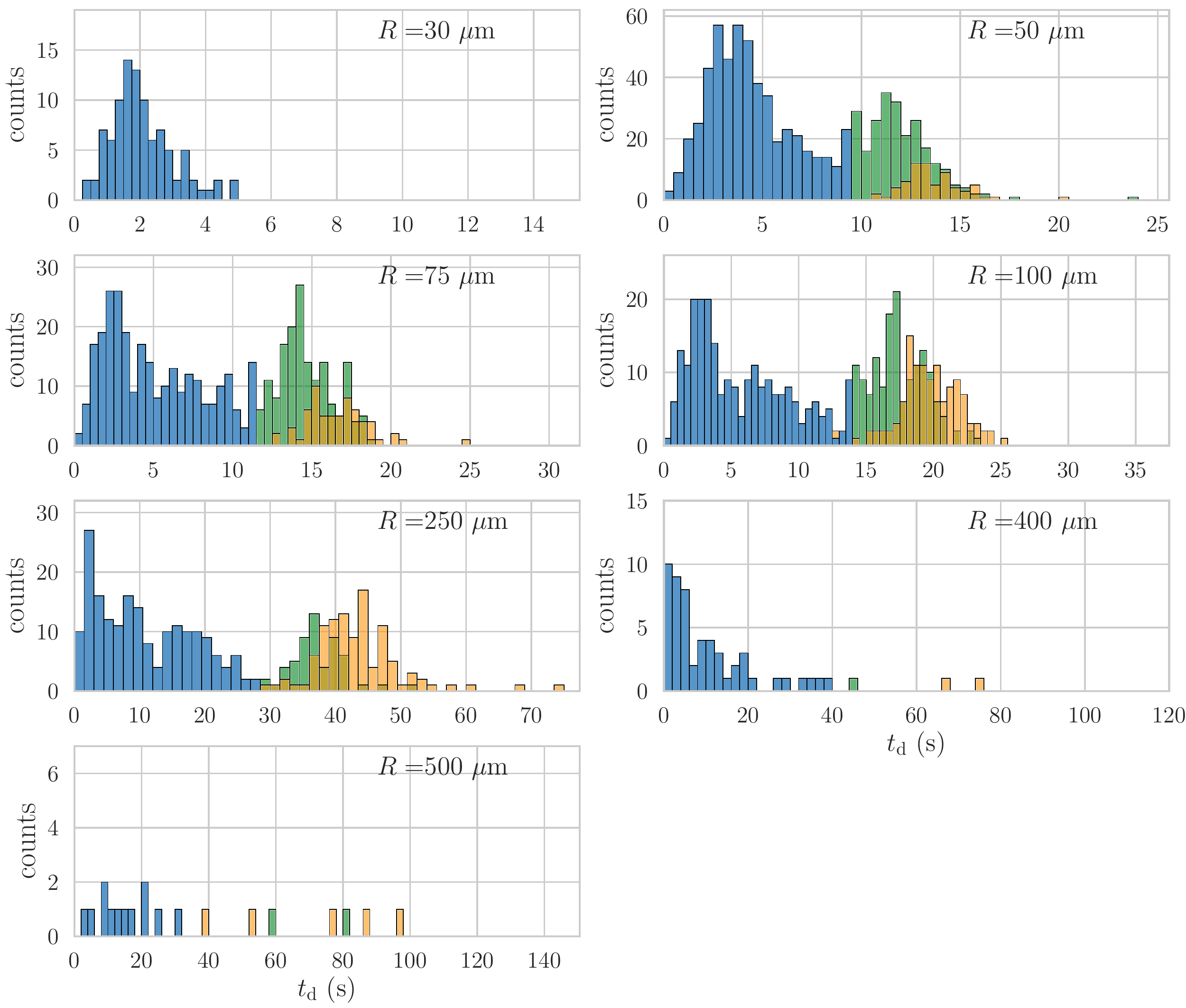}
  \captionof{figure}{Histograms of the detention times at pillars of varying radius. The data set has been filtered to include only interactions with ``clean'' pillars.}
  \label{SIfig:hist-first-time}
\end{center}\end{minipage}

\subsection{Speed and angular velocity}
We calculate the droplet speed $v$ via the displacement between frames, and the angular velocity $\omega$ by the change in polar angle $\alpha$, using the distance vector $\vec{r}$ to the pillar center. Each trajectory can be segmented into orbits of rank $n$ by $\alpha(t_n)-\alpha(t_{n-1})=2\pi$.
We have plotted distributions of $v$ and $\omega$ by pillar size and orbit rank in \figref{SIfig:speeds} and \figref{SIfig:omega}.

Since, starting from the second orbit, a droplet swims through partially consumed fuel (filled micelles), $v$ should decrease between consecutive orbits, which tendency can be seen in the respective histograms. We assume this decrease to be small in order to use a constant radial chemical gradient, however, we can use the specific angular velocities ($\omega_1,\omega_2$) for our calculation of torques in the case ($R=$\SI{250}{\um}) where there is sufficient statistics for the second orbit.

\noindent\begin{minipage}{\textwidth}\begin{center}
  \includegraphics[width=\textwidth]{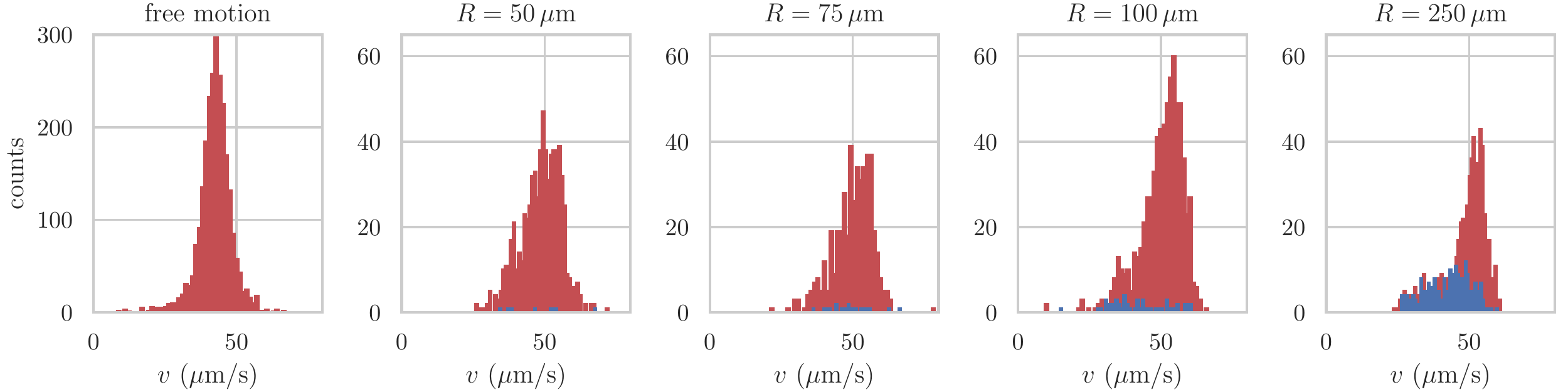}
  \captionof{figure}{Histograms of the segment speeds, first (red) and second (blue) orbit. `Free motion' applies to segments between pillars.}
  \label{SIfig:speeds}
\end{center}\end{minipage}

\noindent\begin{minipage}{\textwidth}\begin{center}
  \includegraphics[width=\textwidth]{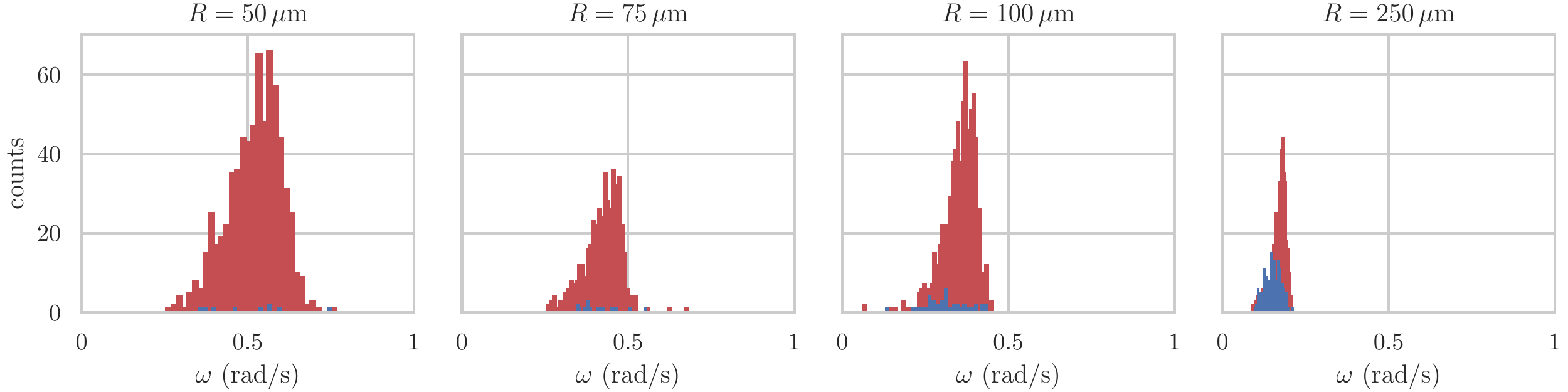}
  \captionof{figure}{Histograms of the segment angular velocities, first (red) and second (blue) orbit.}
  \label{SIfig:omega}
\end{center}\end{minipage}

\subsection{Solubilization rate}\label{SIsec:solub}
A quantitative estimate of the droplet solubilization rate is relevant to our study in two respects: first, we have to assume a constant droplet size during each interaction, otherwise the droplet speed would change, as well as the hydrodynamic interactions. Second, we can use it to relate the chemotactic torque to the initial average concentration of oil molecules $c_0$ dispersed in the droplet trail.

From previous measurements under higher optical resolution (Fig.~5.1 in~\cite{kruger2016_liquid}) we know that the change in radius over time of a 5CB droplet at room temperature in 7.5wt\% TTAB is on the order of $\mathrm{d}R/\mathrm{d}t\approx\SI{5}{\nano\metre\per\second}$.

Since the maximum orbital period (for $R=$\SI{250}{\um}) is $T_\text{orb}\approx \SI{38.5}{\second}$, the relative change in droplet size per orbit, $\frac{\Delta a}{a}(T_\text{orb})\approx .004$, should be negligible. 
\figref{SIfig:disv} shows, at lower resolution and therefore quite noisy, the time evolution of droplet radii for a few trajectories from the present study. Radii decrease by less then 10\% over \SI{300}{\second}, which is consistent with the values given above.

To calculate the number density of dispersed 5CB oil molecules (density \SI{1}{\gram\per\cubic\cm}, molar weight \SI{250}{\gram\per\mol}), we calculate the volume loss per second of an $a=\SI{50}{\micro\metre}$ droplet and assume the corresponding number of molecules to disperse inside a volume given by $(2a)^2\cdot v\cdot\SI{1}{\second}$.

\noindent\begin{minipage}{\textwidth}\noindent\begin{minipage}{.6\textwidth}
  \includegraphics[width=\textwidth]{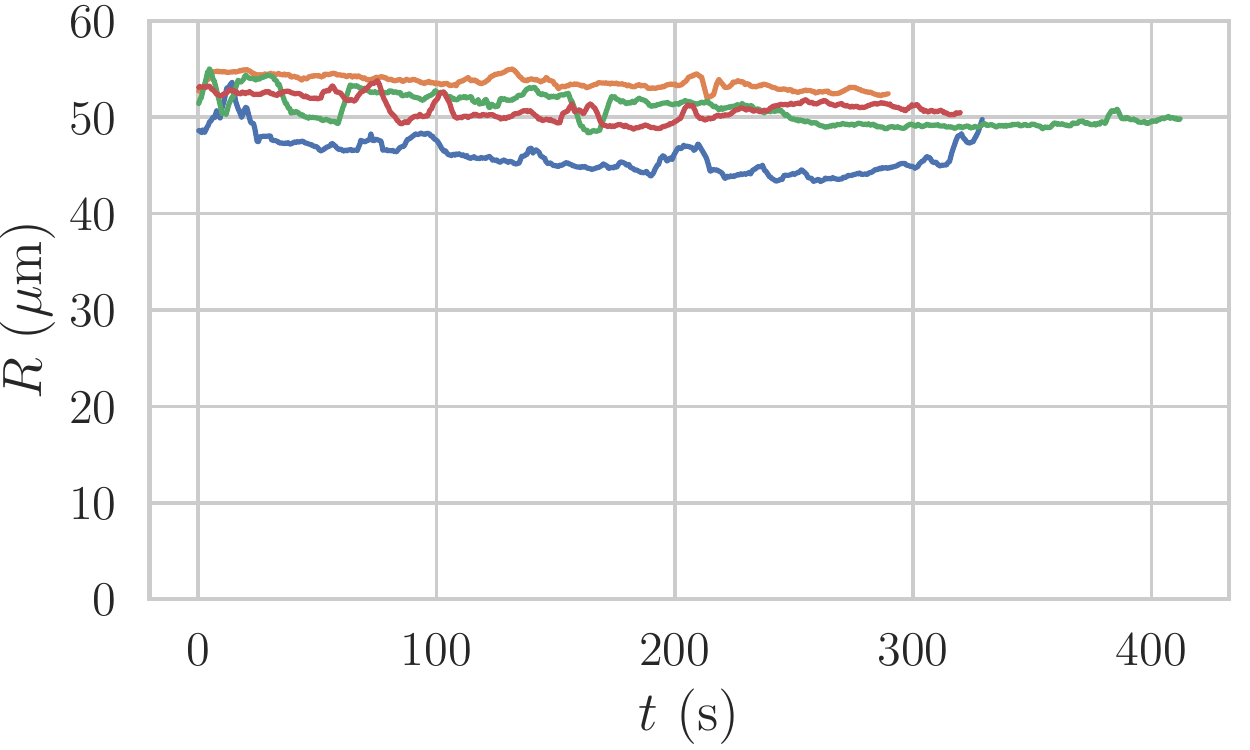}\end{minipage}\noindent\begin{minipage}{.4\textwidth}
  
  \captionof{figure}{Extracted radii (from image binarization) vs. time for several droplets. Droplets do not shrink appreciably during one orbit ($T_\text{orb}<\SI{40}{\second}$).}
  \label{SIfig:disv}
\end{minipage}\end{minipage}

\subsection{Rotational diffusion coefficient}\label{SIsec:diffusion}
\noindent\begin{minipage}{\textwidth}\begin{center}
  \centering
  \includegraphics[width=.8\textwidth]{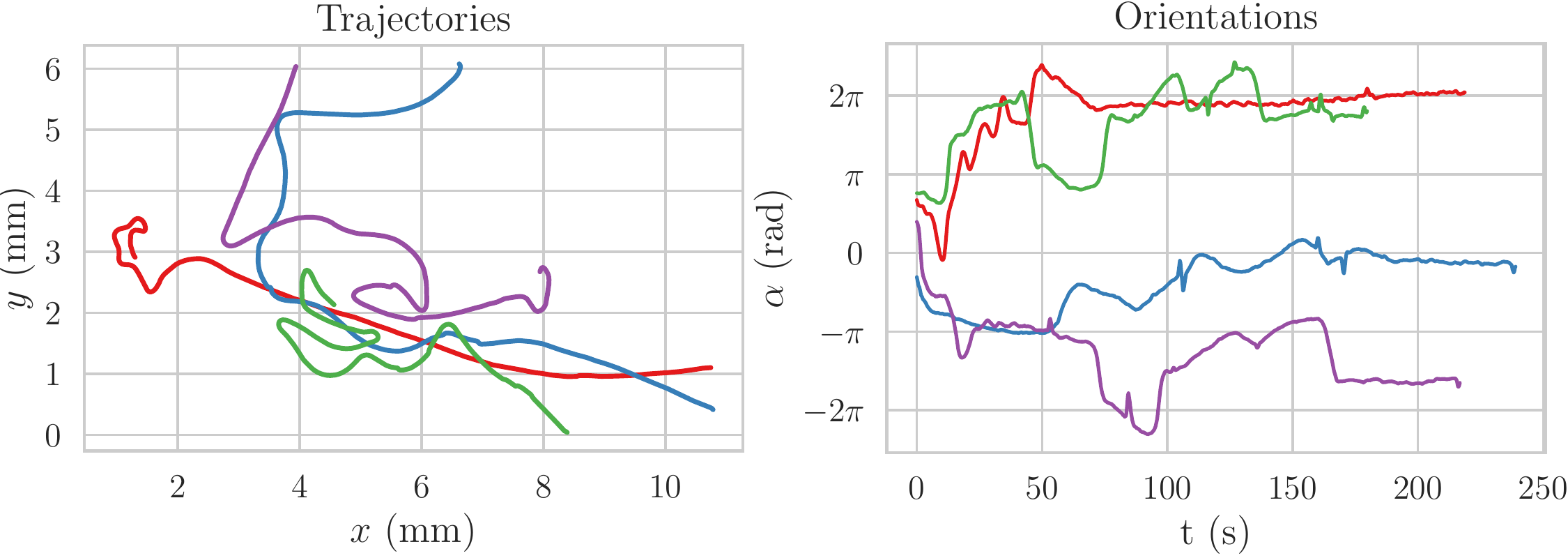}
  \captionof{figure}{Four sample trajectories $\vec{r}$ of swimmers moving between pillars and their orientation $\alpha (t)$}
  \label{SIfig:traj}
\end{center}\end{minipage}

The rotational diffusion coefficient ($D_\text{R}$) can be computed in different ways. 
Following Howse et al. \cite{howse2007_self-motile}, the mean squared displacement of an active Brownian particle is
$$
\text{MSD} =
\begin{cases}
4D_\text{T} \Delta t + v^2 \Delta t^2 \hspace{.8cm} \text{for} \hspace{.4cm} t \ll \tau_\text{R}\\
(4D_\text{T} + v^2 \tau_R) \Delta t \hspace{.8cm} \text{for} \hspace{.4cm} t \gg \tau_R
\end{cases},
$$
where $\tau_R = 1/D_\text{R}$.

In the case of our comparatively large droplets ($a=\SI{50}{\um}$), we can neglect the thermal diffusion $D_\text{T}$. In principle, $\tau_r$ can be determined by 
by analysing the long-time dynamics of the swimmers, calculating the MSD and identifying either the crossover time scale or fitting to diffusive limit. This is, however, not a good approach for persistent swimmers with small $D_\text{R}$, which within our accessible field of view (\SI{10}{\mm} $\times$ \SI{6}{\mm}) do not exhibit a full diffusive crossover (\figref{SIfig:msqr}).

We can, however, estimate $\tau_R$ by analyzing the decay of the angular correlation, $C(t)=\langle \vec{e}(\tau)\vec{e}\cdot(\tau+t)\rangle_\tau$~\cite{saragosti2012_modeling}. Inspection of the correlation decay plotted in \figref{SIfig:msqr} (a) for 4 trajectories shows that $\tau_R$ is on the order of \SI{15}{\second} to \SI{70}{\second}, corresponding to a $D_\text{R}$ between \SI{0.015}{\radian\per\second} and \SI{0.067}{\radian\per\second}.

We can also calculate $D_\text{R}$ via the mean square rotation of the swimmer by $\langle |\Delta \alpha|^2 \rangle = 2 D_\text{R} \Delta t$ \cite{drescher2011_fluid}. 
In the left plot of \figref{SIfig:msqr}, we calculate $D_\text{r}$ from a linear fit to $\langle |\Delta \alpha|^2 \rangle$ for $t<\SI{25}{\second}$ (solid lines) for the same set of trajectories, resulting in $D_\text{R}$ values of $0.041$, $0.063$, $0.090$, $0.015$ respectively.
Thus, we assume $D_\text{R}$  for droplets swimming off-wall to be on the order of \SI{0.05}{\radian\per\second}.

\noindent\begin{minipage}{\textwidth}\begin{center}
  \includegraphics[width=\textwidth]{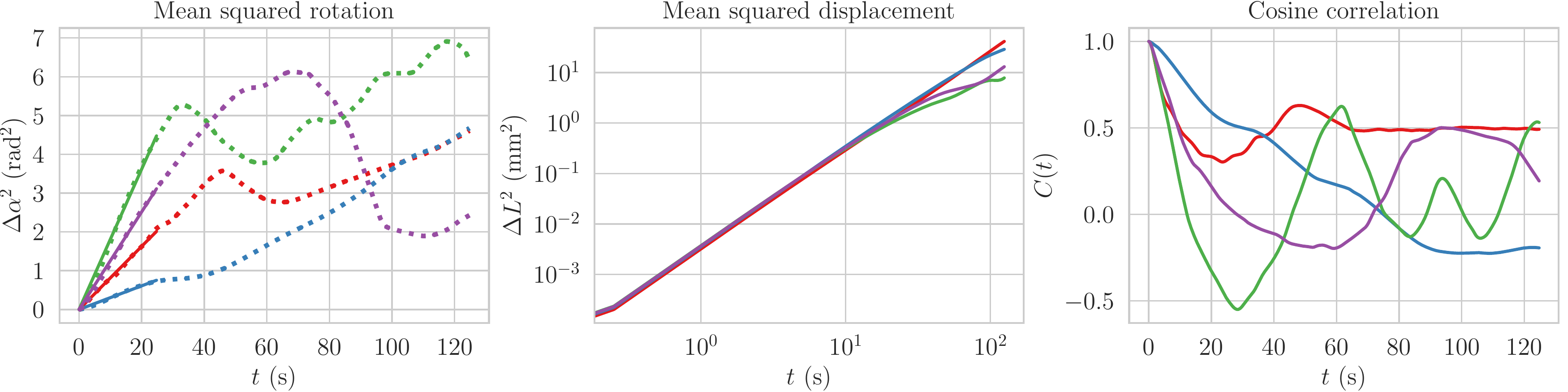}
  \captionof{figure}{Mean squared rotation, mean squared displacement and cosine correlation for 4 trajectories of droplets moving between pillars, as used in the estimation of the rotational diffusion coefficient.}
  \label{SIfig:msqr}
\end{center}\end{minipage}

\clearpage

 \subsection{Data tables for the plots in \figref{fig:analysis}, MS}\label{SIsec:analysis}
 
 \begin{tabular*}{\columnwidth}{@{\extracolsep{\fill}}l|cccc}
    \hline
    $R$ (\si{\um}) & $\langle t_\text{d1} \rangle$ (\si{\s}) & $D_\text{R1}$ (\si{\radian\squared\per\second}) 
    & $\langle t_\text{d2} \rangle$ (\si{\s})  & $D_\text{R2}$ (\si{\radian\squared\per\second}) \\
    \hline
50 & $9.3\pm 0.8$ & $0.05\pm 0.02$ & $1.05\pm 0.06$ & $0.12\pm 0.05$ \\
75 & $46\pm 30$ & $0.08\pm 0.03$ & $1.46\pm 0.07$ & $0.12\pm 0.05$ \\
100 &($2.82\times 10^8$) & $0.07\pm 0.03$ & $1.64\pm 0.05$ & $0.07\pm 0.03$ \\
250 &$(4.25\times 10^8$) & $0.05\pm 0.03$ & $5.4\pm 0.2$ & $0.02\pm 0.01$ \\
\hline
  \end{tabular*}\vspace{1em}
  
  \begin{tabular*}{\columnwidth}{@{\extracolsep{\fill}}l|ccccc}
    \hline
    $R$ (\si{\um}) & $|\omega_1|$ (\si{\radian\per\second}) & $|\omega_2|$ (\si{\radian\per\second}) & $\Omega_\text{W}$ & $\Omega_\text{C}$ & $\partial_r C$ \\
    \hline
50 & $0.50\pm 0.08$ & $0.49\pm 0.12$ & $0.39\pm 0.09$ & $-0.82\pm 0.23$ & -0.0054 \\
75 & $0.41\pm 0.06$ & $0.40\pm 0.05$ & $0.38\pm 0.06$ & $-0.64\pm 0.16$ & -0.0051 \\
100 & $0.34\pm 0.05$ & $0.30\pm 0.06$ & $0.34\pm 0.05$ & $-0.62\pm 0.15$ & -0.0048 \\
250 & $0.16\pm 0.02$ & $0.15\pm 0.02$ & $0.16\pm 0.02$ & $-0.20\pm 0.06$ & -0.0029 \\
    \hline
  \end{tabular*}
 
 \section{Error estimates}
 In \figref{fig:analysis}, we provide error estimates for the quantities $\langle t_d\rangle$, $\omega$, $\Omega_\text{W}$, $\Omega_\text{C}$  and $D_r$. 
 We directly estimate errors from experimental statistics and fitting procedures for 
 $\theta_\text{e}$, $\omega$ and the fitting parameters $\lambda$ and $\mu$, and infer the values for 
 $\langle t_d\rangle$, $\Omega_\text{W}$, $\Omega_\text{C}$  and $D_r$ via the relations 
 $\lambda = \theta_\text{e}^2/2D_\text{R}$ and $\mu = \theta_\text{e} / (\Omega_\text{W}+\Omega_\text{C}-\omega)$ 
 using common rules of error propagation,
 \[\delta Y=\sqrt{\sum_{i=0}^n\left(\frac{\partial Y}{\partial x_i}\delta x_i\right)^2} \text{ for } Y=Y(x_0,\ldots,x_n)\]
 We note that for $R\in\{100,250\}\SI{}{\um}$, $\langle t_d\rangle$ and the respective error diverge for the first orbit (experimentally, we cannot distinguish between inverse Gaussian and L\'evy distributions). In these cases, we have not added error bars.

\section{Numerical gradient estimation}\label{SIsec:gradient}
\raggedright To illustrate the evolution of the chemical gradient at the pillar wall, we have plotted the numerical solution of \eqnref{eqn:main} for the pillar sizes discussed in the main MS in \figref{SIfig:numericconc}.
\noindent\begin{minipage}{\textwidth}\begin{center}
    \includegraphics[width=\textwidth]{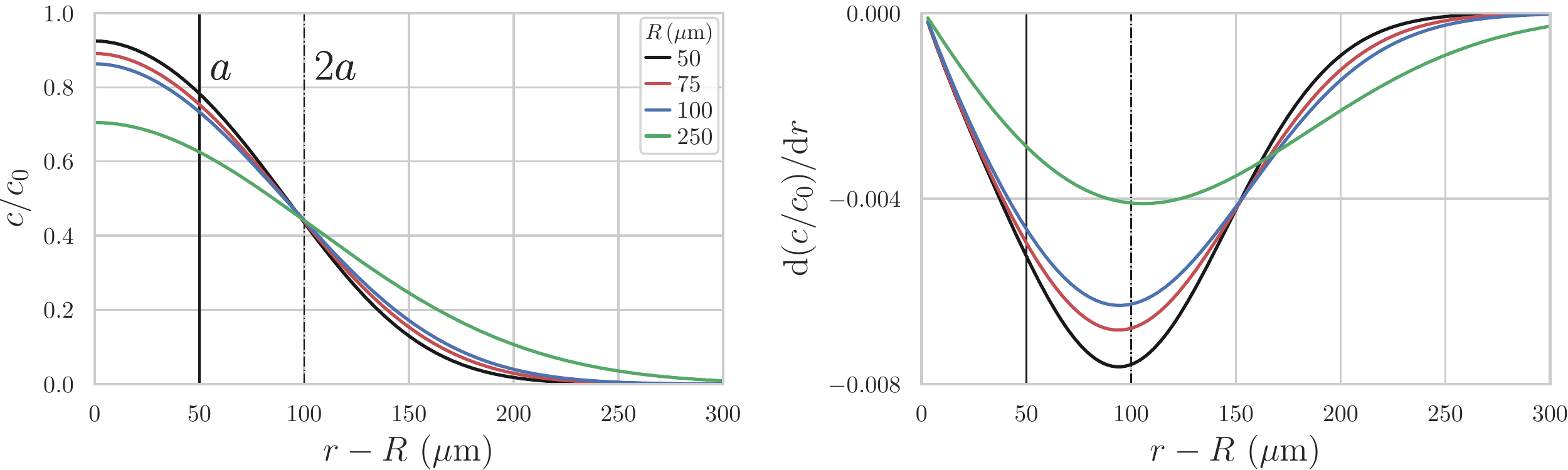}
    \captionof{figure}{(a) Numerically calculated radial concentration  profile (normalised to initial concentration $c_0$) of trails around different pillars at $t=T_\text{orb}$, starting at the pillar wall ($r=R$). Solid and dashed vertical lines indicate the center and outer edge of a droplet of radius $a=$\SI{50}{\um}) attached to the wall.  (b) Rescaled $d(c/c_0)/dr$ at $t=T_\text{orb}$. }
    \label{SIfig:numericconc}
\end{center}\end{minipage}

\section{Fokker-Planck equation for the distribution of detention times at straight walls}\label{SIsec:FP}

\noindent\begin{minipage}{\textwidth}\begin{center}
  \includegraphics[width=.5\textwidth]{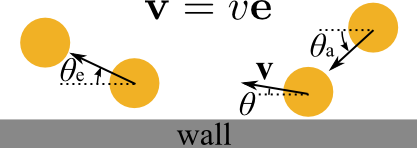}
  \captionof{figure}{Illustration of the swimmer interacting with a straight wall.}
  \label{SIfig:coord}
\end{center}\end{minipage}
Here we include a Fokker-Planck solution of a swimmer interacting with a flat wall with an optional reflective boundary condition to estimate whether this condition is safe to ignore in our more approximative Langevin model.

We model our swimmer with an active Brownian particle moving at a constant speed $v$.
The swimming direction is described by an overdamped Langevin equation
\begin{equation*}
    \frac{d \mathbf{e}}{dt} = \boldsymbol{\Xi}\times\mathbf{e}, \, \text{with } \boldsymbol{\Xi} = \boldsymbol{\Omega} + \mathbf{\xi}\,.
\end{equation*}

From this equation we can  a Fokker-Planck equation that accounts for the evolution in time of the one-particle probability density function, $P(\mathbf{e}, t|\mathbf{e_0}, t_0)$, of finding an active particle, with the initial condition $P(\mathbf{e}, t = t_0|\mathbf{e_0}, t_0) = \delta(\mathbf{e}-\mathbf{e_0})$. 
In the following, to lighten the notation we will use $P(\mathbf{e},t) = P(\mathbf{e},t|\mathbf{e_0},t_0)$. To derive the Fokker-Planck equation, we consider the derivative of $P(\mathbf{e}, t) = \langle\delta (\mathbf{e}-\mathbf{e_0})\rangle$ with respect to time \cite{konotop1994_nonlinear}
\[
\partial_t P(\mathbf{e}, t) = -\sum\limits_{i=1}^n \partial_{e_i}\langle \delta (\mathbf{e}-\mathbf{e_0})\frac{d e_i}{dt}\rangle\,.
\]
The time derivative of $\langle\delta (\mathbf{e}-\mathbf{e_0})\rangle$ reads
\begin{align*}
\partial_t \langle\delta (\mathbf{e}-\mathbf{e_0})\rangle &= \langle\partial_t  \delta (\mathbf{e}-\mathbf{e_0})\rangle\\
&=-\nabla_e\langle \frac{d\mathbf{e}}{dt}\delta(\mathbf{e}-\mathbf{e_0})\rangle\\
&=-\nabla_e\langle (\boldsymbol{\Xi}\times \mathbf{e})\delta(\mathbf{e}-\mathbf{e_0})\rangle\\
&=-\nabla_e\langle (\boldsymbol{\Omega}\times \mathbf{e})\delta(\mathbf{e}-\mathbf{e_0})\rangle -\nabla_e\langle (\mathbf{\xi}\times \mathbf{e})\delta(\mathbf{e}-\mathbf{e_0})\rangle
\end{align*}
By using the Furutsu-Novikov-Donsker relation \cite{sevilla2015_smoluchowski,sevilla2014_theory,kampen1992_stochastic,klyatskin2014_stochastic,konotop1994_nonlinear,frank2005_delay,novikov1965_functionals,zinn-justin2002_quantum,vachier2019_dynamics}
\[
\langle \xi(t) R[\xi]\rangle = \int\limits_{-\infty}^{+\infty}dt'\langle \xi(t)\xi(t')\rangle = \langle \frac{\delta R[\xi]}{\delta\xi(t')}\rangle\,,
\]
the Fokker-Planck equation reads
\begin{equation}
    \partial_t P(\mathbf{e},t)=D_\text{R} \text{L}_e^2P(\mathbf{e},t)-\nabla_e\cdot (\boldsymbol{\Omega}\times \mathbf{e})P(\mathbf{e},t)\,,
    \label{SIeqn:FPE}
\end{equation}
where $\text{L}_e^2$ is the angular part of the Laplacian. Let us take a step back and assume the drift $\Omega$ be null. Then we can rewrite \eqnref{SIeqn:FPE} as
\[
\partial_t P(\mathbf{e},t) = O_p P(\mathbf{e},t)\,.
\]
This system can be then mapped onto the quantum mechanical problem of the rigid rotator and by using a path integral representation \cite{zinn-justin2002_quantum} it is possible to find the solution of the Fokker-Planck equation $P(\mathbf{e},t)$ in $d$ dimensions. By using the bracket notation $P(\mathbf{e},t)$ reads
\[
P(\mathbf{e},t) = \langle \mathbf{e}|e^{D_\text{R} L^2 t}|\mathbf{e_0}\rangle\,.
\]
From \cite{zinn-justin2002_quantum}, the path integral representation for the probability distribution $P(\mathbf{e},t)$ in a free 2D plane can be found and reads 
\[
P(\theta,t|\theta_0,t_0) = \frac{1}{\sqrt{4\pi D_\text{R} t}} \exp(-\frac{(\theta-\theta_0)^2}{4 D_\text{R} t})\,.
\]

In order to calculate the distribution of detention times, we need to impose an absorbing boundary condition at the escape angle $\theta_\text{e}$ , $[P(\theta_\text{e},t)=0]$. By using the method of images, the probability distribution can be found and reads
\[
P(\theta,t|\theta_0, t_0) = \frac{1}{\sqrt{4\pi D_\text{R} t}} \left[ \exp(-\frac{(\theta-\theta_0)^2}{4 D_\text{R} t}) -\exp(-\frac{(\theta-(2\theta_\text{e} -\theta_0))^2}{4 D_\text{R} t}) \right] \,.
\]

The survival probability is given by
\begin{align*}
S(t)&=\int\limits_{-\infty}^{\theta_\text{e}}P(\theta,t|\theta_0, \theta_\text{e})d\theta \\
&=\frac{1}{2}\left[\erf(\frac{\theta_\text{e}-\theta_0}{\sqrt{4D_\text{R} t}})+1-\erf(\frac{\theta_0-\theta_\text{e}}{\sqrt{4D_\text{R} t}})-1\right]\\
&=\frac{1}{2}\left[\erf(\frac{\theta_\text{e}-\theta_0}{\sqrt{4D_\text{R} t}})+\erf(\frac{\theta_\text{e}-\theta_0}{\sqrt{4D_\text{R} t}})\right]\\
&=\erf(\frac{\theta_\text{e}-\theta_0}{\sqrt{4D_\text{R} t}}) \,.
\end{align*}

The detention time follows a L\'evy distribution
\begin{align}
f(t)&=-\frac{\partial}{\partial t}S(t)\nonumber\\
&=\frac{(\theta_\text{e}-\theta_0)}{\sqrt{4\pi D_\text{R} t^3}} \exp({-\frac{(\theta_\text{e}-\theta_0)^2}{4 D_\text{R} t}}) \,.
\label{SIeqn:levy}
\end{align}

\noindent\begin{minipage}{\textwidth}\begin{center}
    \centering
    \includegraphics[width=.6\textwidth]{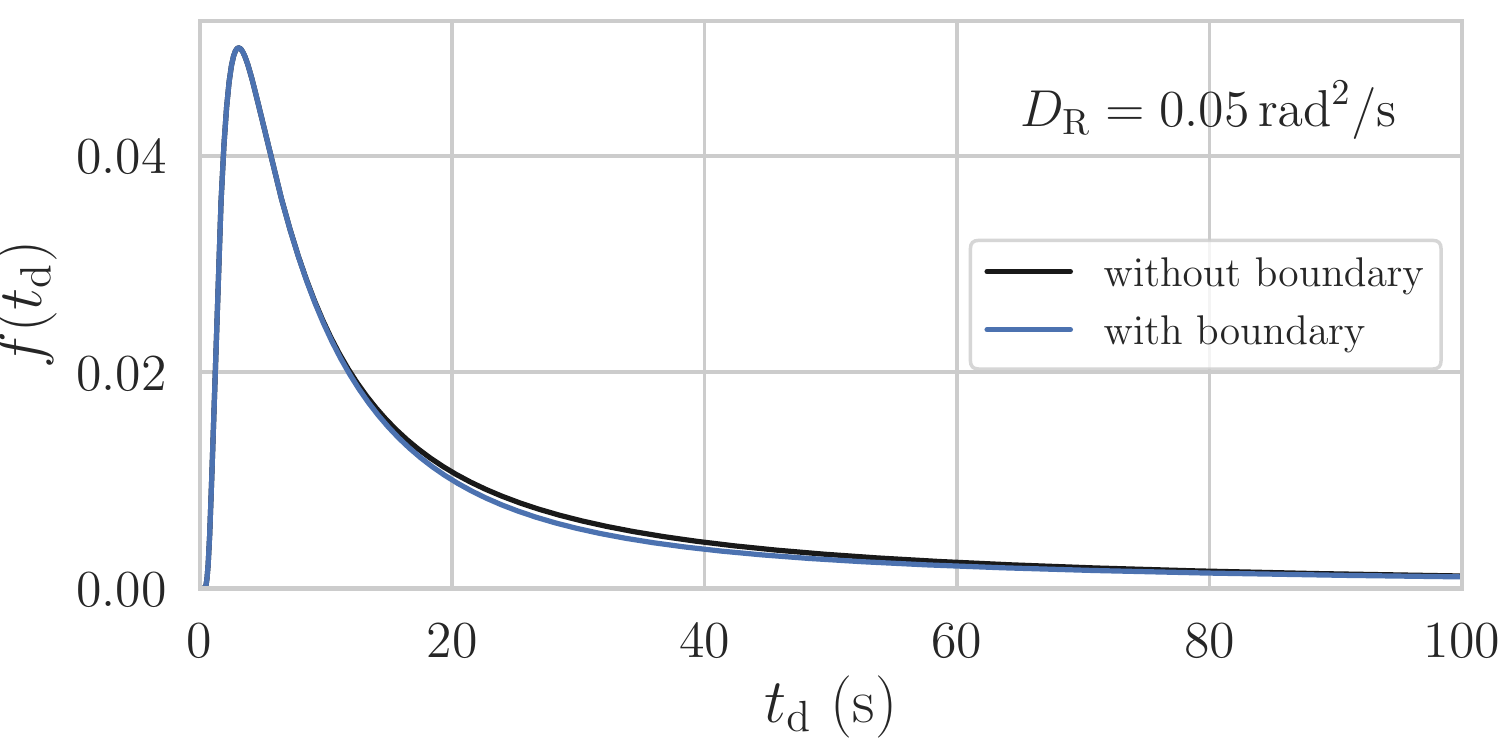}
    \captionof{figure}{L\'evy distribution of the first passage time for a Brownian process without drift, with (\eqnref{SIeqn:levy-b}, blue curve) and without (\eqnref{SIeqn:levy}, black curve) the reflective boundary condition. For this plot we take $\theta_0 =0$, $\theta_\text{e} = -0.962$, and $\theta_\text{r} = \pi$.}
    \label{SIfig:levy}
\end{center}\end{minipage}

We now apply a reflective boundary condition $[\partial_{\theta}P(\theta,t)|_{\theta_\text{r}}=0]$.
By using the method of images, the survival probability is given by

\begin{align*}
   S(t)=&\int\limits_{\theta_\text{r}}^{\theta_\text{e}}P(\theta,t|\theta_0,t_0)d\theta\\
  =&\frac{1}{2}\left[ \erf(\frac{\theta_0-\theta_\text{r}}{\sqrt{4D_\text{R} t}})-\erf(\frac{\theta_0-\theta_\text{e}}{\sqrt{4D_\text{R} t}})  -\erf(\frac{\theta_0-\theta_\text{e}}{\sqrt{4D_\text{R} t}})+\erf(\frac{\theta_0-2\theta_\text{e}+\theta_\text{r}}{\sqrt{4D_\text{R} t}})\right.\\
  &\left. +\erf(\frac{\theta_0+\theta_\text{e}-2\theta_\text{r}}{\sqrt{4D_\text{R} t}})-\erf(\frac{\theta_0-2\theta_\text{e}+\theta_\text{r}}{\sqrt{4D_\text{R} t}}) \right]\,. 
\end{align*}

The detention time is given by $f(t)=-\frac{\partial}{\partial t}S(t)$ and reads

\begin{align}
    f(t) =& -\frac{1}{2\sqrt{4\pi D_\text{R} t^3}}\left[(\theta_0-\theta_\text{r})\exp(-\frac{(\theta_0-\theta_\text{r})^2}{4D_\text{R} t}) -(\theta_0-\theta_\text{e}) \exp(-\frac{(\theta_0-\theta_\text{e})^2}{4D_\text{R} t}) \right.\nonumber\\
    &\left. -(\theta_0-\theta_\text{e}) \exp(-\frac{(\theta_0-\theta_\text{e})^2}{4D_\text{R} t})
    +(\theta_0 -2\theta_\text{e} + \theta_\text{r}) \exp(-\frac{(\theta_0 -2\theta_\text{e} + \theta_\text{r})^2}{4D_\text{R} t})\right.\nonumber\\
    &\left. +(\theta_0+\theta_\text{e}-2\theta_\text{r})\exp(-\frac{(\theta_0+\theta_\text{e}-2\theta_\text{r})^2}{4D_\text{R} t}) 
    -(\theta_0-\theta_\text{r}) \exp(-\frac{(\theta_0-\theta_\text{r})^2}{4D_\text{R} t})\right] \,.
    \label{SIeqn:levy-b}
\end{align}

We plot \eqnref{SIeqn:levy} and \eqnref{SIeqn:levy-b} with parameters relevant to our experimental system ($\theta_0 =0$, $\theta_\text{e} = -0.962$, $\theta_\text{r} = \pi$, and $D_\text{R}=0.05 \,\text{rad}^2/\text{s}$) in \figref{SIfig:levy}; the difference between two curves is indeed negligible. 

We note that for different $D_\text{R}$, the distribution curves will be similar in shape, as \eqnref{SIeqn:levy} can be written as
$$
f(t) = D_\text{R}\frac{(\theta_\text{e}-\theta_0) }{\sqrt{4\pi \Bar{t}^3}} \exp({-\frac{(\theta_\text{e}-\theta_0)^2}{4 \Bar{t}}}) \,.
$$
with $\Bar{t} = D_\text{R}t$. This also holds for \eqnref{SIeqn:levy-b}.

\section{Detention time distribution at large pillars}
The detention time distributions at large pillars have a large mean, such that $f(t_d)$ is close to the L\'evy distribution for a flat wall, as derived above.
In \figref{SIfig:hist-other} we show the normalised detention time histograms at large pillars (radii \SI{400}{\um} and \SI{500}{\um}), and fits with both inverse Gaussian and L\'evy distributions. The differences are within the experimental error.

\noindent\begin{minipage}{\textwidth}\begin{center}
  \includegraphics[width=.85\textwidth]{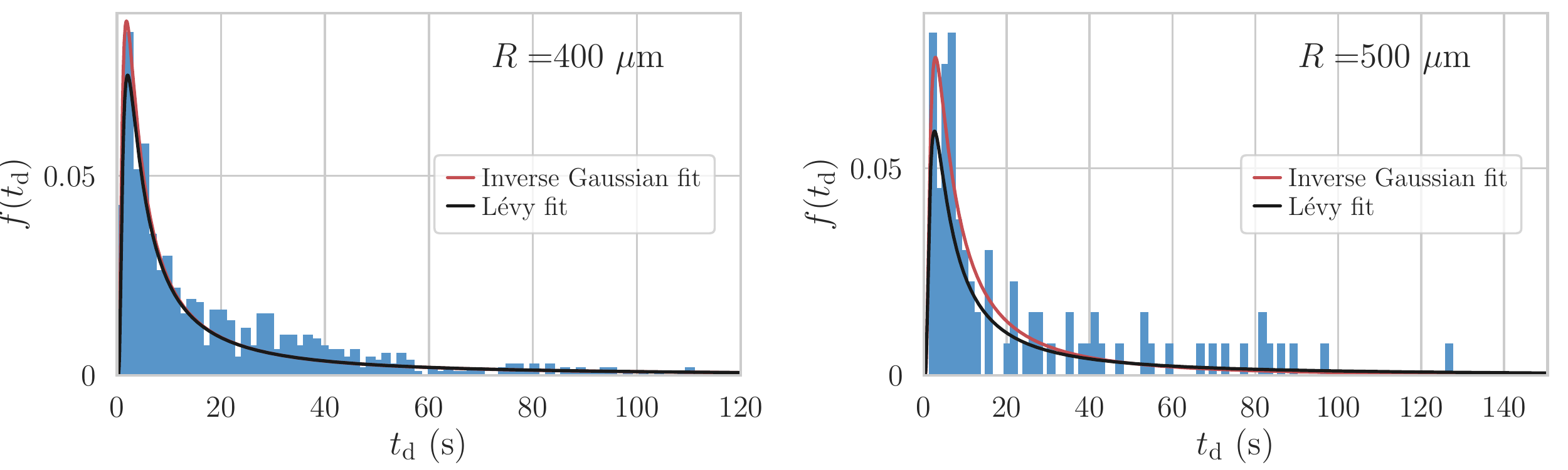}
  \captionof{figure}{Normalised histograms of the detention times at large pillars (radius \SI{400}{\um} and \SI{500}{\um}). using data from all pillar interactions. Solid lines correspond to fits to inverse Gaussian (red) and L\'evy distributions (black).}
  \label{SIfig:hist-other}
\end{center}\end{minipage}

\section{Supporting movie}\label{SIsec:movie}
\noindent\begin{minipage}{\textwidth}\noindent\begin{minipage}{.5\textwidth}
  \includegraphics[width=.8\textwidth]{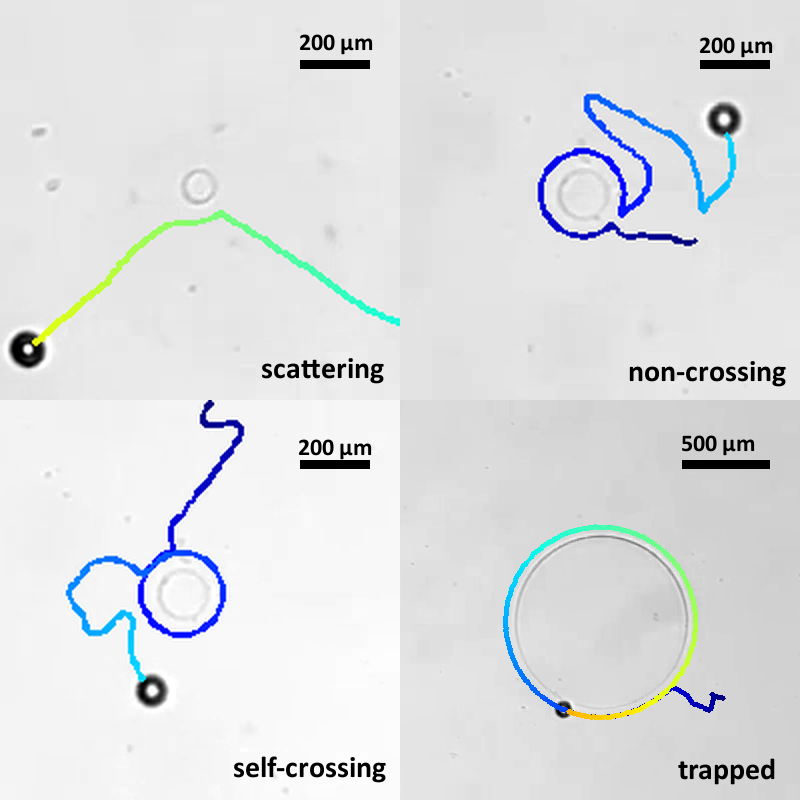}\hspace{.03\textwidth}\includegraphics[width=.155\textwidth]{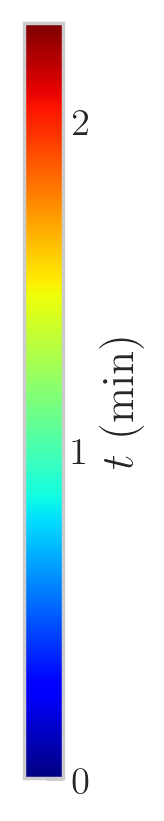}\end{minipage}\hspace{.05\textwidth}\noindent\begin{minipage}{.45\textwidth}
  
  \captionof{figure}{Supporting movie, with examples of scattering ($R=\SI{50}{\um}$), non-crossing ($R=\SI{75}{\um}$), self-crossing ($R=\SI{75}{\um}$) and trapping  ($R=\SI{500}{\um}$) interactions. The raw movies were recorded at 4 frames/second and have been sped up by a factor of 6. The trajectory colour marks the time in the individual experiment.}
  \label{SIfig:movie}
\end{minipage}\end{minipage}
\end{document}